\documentclass[11pt,a4paper]{article}
\pdfoutput=1
\usepackage{jheppub}

\usepackage[utf8]{inputenc}
\usepackage[dvipsnames]{xcolor}
\usepackage{verbatim}
\usepackage{textcomp}
\usepackage{amsmath}
\usepackage{amssymb}
\usepackage{graphicx}
\usepackage{nicefrac,mathrsfs}

\usepackage{booktabs,footnote,multirow}
\usepackage{fmdg}
\usepackage{array,bm}
\usepackage{multirow}
\usepackage{tikz}
\usepackage[compat=1.1.0]{tikz-feynman}
\usepackage{subcaption}
\usepackage{caption}
\usepackage{tikz}
\usepackage[compat=1.1.0]{tikz-feynman}
\usetikzlibrary{calc}
\usetikzlibrary{shapes.misc}
\tikzset{
	>=latex,
    photon/.style={decorate, decoration={snake}, draw=black, thick},
    fermionnoarrow/.style={draw=black, postaction={decorate}, thick},
    scalar/.style={draw=black, postaction={decorate}, decoration={markings,mark=at position .55 with {\arrow{>}}}, thick, dashed},
    scalarnoarrow/.style={draw=black, postaction={decorate},  thick, dashed},
    fermion/.style={draw=black, postaction={decorate},decoration={markings,mark=at position .55 with {\arrow{>}}}, thick},
    gluon/.style={decorate, draw=black, decoration={coil,amplitude=4pt, segment length=5pt}, thick},
    vertex/.style={draw,shape=circle,fill=black,minimum size=3pt,inner sep=0pt},
    fillvertex/.style={draw,shape=circle,fill=black,minimum size=5pt,inner sep=0pt},
    openvertex/.style={draw,shape=circle,minimum size=51pt,inner sep=0pt},
    blob/.style={draw=black,shape=circle,fill=black,minimum size=6pt,inner sep=0pt},
    redvertex/.style={draw=red,shape=circle,fill=red,minimum size=3pt,inner sep=0pt},
    cross/.style={cross out, draw=black,thick, minimum size=5pt, inner sep=0pt, outer sep=0pt}
}

\usepackage{slashed,gensymb}
\usepackage{cancel,empheq}

\definecolor{MS}{rgb}{1,0,0}                                                                   
\definecolor{all}{rgb}{1,0,1}

\newcommand{\calM}{\mathcal{M}}
\newcommand{\calA}{\mathcal{A}}
\newcommand{\calO}{\mathcal{O}}
\newcommand{\calD}{\mathcal{D}}

\newcommand{\phiR}{\phi_R}

\title{The ScotoSinglet Model: A Scalar Singlet Extension of the Scotogenic Model}

\author[a]{Ankit Beniwal,}
\author[b]{Juan Herrero-Garc\'ia,}
\author[c]{Nicholas Leerdam,}
\author[c]{Martin White}
\author[c]{and Anthony G. Williams}

\affiliation[a]{Center for Cosmology, Particle Physics and Phenomenology (CP3), \\ Universit\'{e} catholique de Louvain, B-1348 Louvain-la-Neuve, Belgium}
\affiliation[b]{Departamento de F\'isica Te\'orica and IFIC, Universidad de Valencia-CSIC, \\ C/ Catedr\'atico Jos\'e Beltr\'an, 2, E-46980 Paterna, Spain}
\affiliation[c]{ARC Centre of Excellence for Dark Matter Particle Physics and CSSM, \\ Department of Physics, University of Adelaide, SA 5005, Australia}

\emailAdd{ankit.beniwal@uclouvain.be}
\emailAdd{juan.herrero@ific.uv.es}
\emailAdd{nicholas.leerdam@adelaide.edu.au}
\emailAdd{martin.white@adelaide.edu.au}
\emailAdd{anthony.williams@adelaide.edu.au}

\abstract{The Scotogenic Model is one of the most minimal models to account for both neutrino masses and dark matter (DM). In this model, neutrino masses are generated at the one-loop level, and in principle, both the lightest fermion singlet and the lightest neutral component of the scalar doublet can be viable DM candidates.~However, the correct DM relic abundance can only be obtained in somewhat small regions of the parameter space, as there are strong constraints stemming from lepton flavour violation, neutrino masses, electroweak precision tests and direct detection.~For the case of scalar DM, a sufficiently large lepton-number-violating coupling is required, whereas for fermionic DM, coannihilations are typically necessary. In this work, we study how the new scalar singlet modifies the phenomenology of the Scotogenic Model, particularly in the case of scalar DM. We find that the new singlet modifies both the phenomenology of neutrino masses and scalar DM, and opens up a large portion of the parameter space of the original model.}

\preprint{ADP-20-25/T1135, CP3-20-46, IFIC/20-46}


\arxivnumber{2010.05937}

\makeatother

\begin{document}

\maketitle

\section{Introduction} \label{sec:intro}
The origin of light neutrino masses and dark matter (DM) remain unsolved puzzles of the Standard Model (SM). Many explanations have been proposed; of particular interest are those that can simultaneously account for both of them.~A successful model should not only provide a natural explanation for the smallness of neutrino masses, it should also include a viable DM candidate.~For the former case, radiative neutrino mass models (see refs.~\cite{Boucenna:2014zba,Cai:2017jrq} for a review), with new states at the TeV scale, are an attractive solution. As for DM, a GeV--TeV scale Weakly Interacting Massive Particle, stabilised by some remanent symmetry (e.g., $\mathbb{Z}_2$), provides an elegant solution.~In this context, the Scotogenic Model (ScM), first proposed by Ernest Ma (2006) \cite{Ma:2006km} (see also refs.~\cite{Kubo:2006yx,Sierra:2008wj,Gelmini:2009xd,Suematsu:2009ww,Hambye:2009pw,Schmidt:2012yg,Racker:2013lua,Toma:2013zsa,Molinaro:2014lfa,Vicente:2014wga,Ahriche:2016cio,Hessler:2016kwm,Avila:2019hhv,Ahriche:2017iar,Ahriche:2018ger,Baumholzer:2019twf,Sarma:2020msa,Escribano:2020iqq}), emerges as one of the simplest joint solutions for neutrino masses and DM. It adds a new scalar doublet $\Phi$ and new Majorana singlets $\psi_i$ ($i \geq 2$), all of them charged under a $\mathbb{Z}_2$ symmetry.~Another interesting variant has been recently proposed where the $\mathbb{Z}_2$ symmetry is exchanged for a $U(1)$ symmetry; it is referred to as the Generalised Scotogenic Model (GScM)~\cite{Hagedorn:2018spx}.

In the ScM, neutrino masses are generated at the one-loop level, and are proportional to a particular quartic coupling, $\lambda_{H\Phi,3}$, and the Majorana masses, $m_{\psi_i}$.~In principle, DM can either be in the form of the lightest $\psi_i$, or the lightest neutral component of $\Phi$ (either CP-even or CP-odd, depending on the sign of $\lambda_{H\Phi,3}$).~In both cases, however, saturating the observed DM relic abundance is non-trivial as the models are subject to strong constraints from neutrino masses, electroweak precision tests (EWPT), direct detection (DD) experiments, and particularly lepton flavour violation (LFV) processes; the DD limits, however, can be circumvented by imposing a lower bound on the splitting between the CP-even and CP-odd scalars, i.e., a lower bound on $\lambda_{H\Phi,3}$.~For the fermion DM case, coannihilations are typically required \cite{Hagedorn:2018spx}.

In this work, we investigate how the parameter space of the original ScM is augmented in the presence of a real scalar singlet, denoted by $\varphi$.~This can be understood as the simplest extension of the Scotogenic model~\cite{Ma:2006km}, which we refer to as the ScotoSinglet Model (ScSM). Notice that the usual ScM is recovered in the limit of no mixing, and when the singlet decouples.~The model was first outlined in ref.~\cite{Farzan:2009ji}, assuming an MeV-scale scalar DM with annihilations into neutrinos only (see also ref.~\cite{Restrepo:2013aga}, based on the one-loop classification of ref.~\cite{Bonnet:2012kz}).~A scale-invariant version of the ScM with an extra scalar singlet was studied in ref.~\cite{Ahriche:2016cio}.\footnote{In this case, the scalar singlet is not charged under a $\mathbb{Z}_2$ symmetry, and instead plays the role of a dilaton with very different phenomenological implications.}~A similar singlet-doublet model\footnote{A singlet-triplet DM model with radiative neutrino masses is proposed in ref.~\cite{Alcaide:2017xoe}.}, also in the fermion sector, was studied in ref.~\cite{Esch:2018ccs}; for other studies, see refs.~\cite{Cohen:2011ec,Cheung:2013dua,Banik:2014cfa}.~Collider signatures of a similar model (without the $\lambda_5$ term) were studied in ref.~\cite{Ahriche:2020pwq}.~Low-scale leptogenesis was studied in the context of ScM \cite{Hugle:2018qbw} and a real scalar singlet \cite{,Alanne:2018brf}.~Recently, in ref.~\cite{Hashimoto:2020xoz}, inflation was studied in a ScM with an additional scalar singlet (not charged under the $\mathbb{Z}_2$ symmetry). Also recently, a different variant of the ScM with a scalar singlet not charged under $\mathbb{Z}_2$ and spontaneously broken lepton number was studied in ref.~\cite{Bonilla:2019ipe}.~Here we aim to provide a more detailed analysis of the full parameter space while focusing on all possible DM candidates in the ScSM, and comparing our results against the usual ScM. We also study how the parameter space expands with respect to the case of pure singlet or pure doublet DM via turning on/off the relevant couplings.

The ScSM has some interesting features (mostly due to the presence of a scalar singlet-doublet mixing, and a trilinear coupling between $\varphi$, $\Phi$ and the Higgs doublet $H$): new contributions to neutrino masses, 3 potential DM candidates (one of which is a mixture of singlet and doublet), and the possibility to maintain the $\mathbb{Z}_2$ symmetry up to high energy scales. In light of these features, we perform, for the first time, a convergent global fit of the ScSM. As we will show, the presence of the singlet significantly opens up the allowed parameter space of the CP-even scalar DM, which now has a non-negligible singlet component.~Due to the presence of the trilinear coupling with the singlet, there exists a splitting between the CP-even components. This naturally translates into a significant mass splitting between the lightest CP-even and CP-odd scalars, which allows to naturally evade the stringent DD limits.

The rest of the paper is organised as follows. In section~\ref{sec:model}, we introduce the ScSM.\footnote{Our \textsf{FeynRules} \cite{Alloul:2013bka} and \textsf{CalcHEP} \cite{Belyaev:2012qa} model files are available \href{https://feynrules.irmp.ucl.ac.be/wiki/ScotoSinglet}{here}.}~The phenomenology of the ScSM and various theoretical/observational constraints that we impose are described in section~\ref{sec:constraints}.~Sections~\ref{sec:scandet} and \ref{sec:results} are devoted to our numerical analysis and results, respectively.~Our conclusions are presented in section~\ref{sec:conc}.~A list of appendices provide supplementary information for understanding various expressions in the paper.

\section{The ScotoSinglet Model (ScSM)} \label{sec:model}
The new particle fields of the ScSM and their quantum numbers are presented in table~\ref{tab:Scotosinglet}. The Lagrangian for the Majorana fermion fields $\Psi \equiv (\psi_1,\ldots,\psi_N)^T$ is given by
\begin{equation}
	\mathscr{L}_{\Psi} = \frac{1}{2} \overline {\Psi} (i \slashed \partial - M_\Psi) \Psi -\,\overline \Psi \,\mathbf{y_\Psi} \, \tilde\Phi^\dagger\,L\,  \,+\, \text{H.c.}\,, \label{Lpsi}
\end{equation}
where $M_\Psi$ is an $N\times N $ diagonal mass matrix with real and positive values, and $ \mathbf{y_\Psi} $ is an $ N\times 3 $ complex matrix of Yukawa couplings.~Without loss of generality, we take $ m_{\psi_1} \geq m_{\psi_2} \geq \ldots \geq m_{\psi_N} $. To reproduce the neutrino masses, $N \geq 2$. Here we study the minimal case $(N = 2)$.\footnote{For the main purpose of our study (scalar DM and comparison with the ScM), the results are not expected to change significantly for $N> 2$.}

The most general form of a $\mathbb{Z}_2$ symmetric scalar potential is
\begin{align}\label{eqn:V11A}
	V= &- \mu_H^2  H^\dagger H \,+\, \lambda_H (H^\dagger H)^2 \,+\, m_\Phi^2  \Phi^\dagger\Phi\, +\, \lambda_\Phi (\Phi^\dagger\Phi )^2\,+ \frac{1}{2} m^{2}_{\varphi} \varphi^2 \,+\frac{1}{4} \lambda_{\varphi} \varphi^4   \nonumber \\[1mm]
	&\, + \, \lambda_{H\Phi,1} (H^\dagger H) (\Phi^\dagger\Phi)  \,+\, \lambda_{H\Phi,2} (H^\dagger \Phi) (\Phi^\dagger H) \,+\, \frac{1}{2} \left[\lambda_{H\Phi,3} (H^\dagger\Phi)^2 + {\rm H.c.} \right] \nonumber \\[1mm]
	&\,+\frac{1}{2} \lambda_{H\varphi} H^\dagger H \varphi^2  \,+\frac{1}{2} \lambda_{\Phi\varphi} \Phi^\dagger\Phi \varphi^2 \,+\,\left[\kappa\, \Phi^\dagger H \,  \varphi  + {\rm H.c.} \right] \,.
\end{align}
Here $H\equiv \left(0,~(v+h)/\sqrt{2}\,\right)^T$ is the SM Higgs doublet after electroweak symmetry breaking (EWSB).\footnote{The SM Higgs boson mass is $m_h = \sqrt{2\,\lambda_H v^2} = \sqrt{2\,\mu_H^2} = 125$\,GeV, where the vacuum expectation value is $v = \sqrt{\mu^2_H/\lambda_H} = 246.22$\,GeV.}~Without loss of generality, $ \lambda_{H\Phi,3}$ can be made real by performing a rotation of $\Phi$. Although $\kappa$, in general, can be complex, we also take it as real in our study. 

\begin{table}[t]
\centering
	\begin{tabular}{l|cccc}
		\hline \hline
		Fields & $SU(3)_C$ & $SU(2)_L$ & $U(1)_Y$ & $ \mathbb{Z}_2 $ \\ \hline
		Real $\varphi$ & $1$ & $1$ & $0$ & $-$\\ 
		$\Phi$ & $1$ & $2$ & $1/2$ & $-$ \\
		$\psi_k$ & $1$ & $1$ & $0$ & $-$ \\ \hline \hline
	\end{tabular}
	\caption{Particle content of the ScotoSinglet Model (ScSM). Here $k = 1,\ldots,N$ are the number of new Majorana fermion fields; in our study, we consider $N = 2$.}
	\label{tab:Scotosinglet}
\end{table}

We assume that the $\mathbb{Z}_2$ symmetry is exactly preserved such that in the vacuum state, $\langle\Phi\rangle=\langle\varphi\rangle=0$. The electrically neutral and charged field components of the weak scalar doublet $\Phi$ are  
\begin{equation}
	\Phi \equiv 
	\begin{pmatrix}
		\phi^+ \\
		\dfrac{1}{\sqrt{2}}\left(\phi_{R} + i A \right)
	\end{pmatrix}\,.
\end{equation}
After EWSB, the physical masses of charged scalar $\phi^+$ and CP-odd pseudoscalar $A$ are
\begin{subequations}
\begin{align}
	m_{\phi^+}^2 &= m_\Phi^2 + \frac{1}{2} \lambda_{H\Phi,1} v^2 \,, \\[1mm]
	m_A^2 &= m_\Phi^2 + \frac{1}{2}(\lambda_{H\Phi,1} + \lambda_{H\Phi,2} - \lambda_{H\Phi,3}) v^2\,.
\end{align}
\end{subequations}
In the $(\phi_R,\varphi)$ basis, the squared mass matrix is non-diagonal, namely
\begin{align}\label{eqn:nondia_matrix}
	\calM^2 = 
	\begin{pmatrix}
		\dfrac{\partial^2 V}{\partial \phi_R^2} & \dfrac{\partial^2 V}{\partial \phi_R\,\partial \varphi} \\[5mm]
		\dfrac{\partial^2 V}{\partial \varphi\,\partial \phi_R} & \dfrac{\partial^2 V}{\partial \varphi^2} 
	\end{pmatrix}
	=
	\begin{pmatrix}
		 a & c \\
		 c & b
	\end{pmatrix} \,,
\end{align}
where\footnote{Notice that $m_A^2 = a -\lambda_{H\Phi,3} v^2$, where $a$ is the mass of the CP-even scalar doublet in the absence of mixing ($\theta=0$).}
\begin{equation} \label{eq:scmass}
	a = m_\Phi^2 + \frac{1}{2}(\lambda_{H\Phi,1} + \lambda_{H\Phi,2} + \lambda_{H\Phi,3}) v^2  \,, \quad 
	b = m^{2}_{\varphi}+ \frac{1}{2} \lambda_{H\varphi} v^2\,, \quad
	c = \kappa v \,.
\end{equation}
Notice from eq.~\eqref{eqn:V11A} that $\kappa$ controls the mixing between $
\phi_R$ and $\varphi$, and $\lambda_{H\Phi,3}$ the mass splitting  between $\phi_R$ and $A$.~To diagonalise the mass matrix in eq.~\eqref{eqn:nondia_matrix}, we perform a rotation into the physical mass basis $(\eta_1,\,\eta_2)$ by  
\begin{align}\label{eqn:mixingSC}
	\begin{pmatrix}
	\eta_1 \\
	\eta_2
	\end{pmatrix}
	= 
	\begin{pmatrix}
	\cos\theta & \sin\theta \\
	-\sin\theta & \cos\theta \\
	\end{pmatrix}
	\begin{pmatrix}
	\phi_R \\
	\varphi
	\end{pmatrix}\,,
\end{align}
such that
\begin{align} \label{eq:mixing}
	\tan2\theta = \frac{2c}{a-b}\,.
\end{align}
The mixing angle $ \theta \in [0,\,\pi]$ and the quadrant is determined by the signs of $ a-b $ and $ c $. For instance, $ a-b>0 $ and $ c>0 $ implies $ \theta \in (0,\,\pi/4) $, whereas $ a-b<0 $ and $ c>0 $ implies $ \theta \in (\pi/4,\, \pi/2) $. The physical scalar $\eta_{1,2}$ masses are given by
\begin{subequations} \label{eqs:massesab}
	\begin{align}
		m_{\eta_1}^2 &= \frac{1}{2} \left( a+b + \sqrt{(a-b)^2 + 4c^2} \right)\,, \\[1mm]
		m_{\eta_2}^2 &= \frac{1}{2} \left( a+b - \sqrt{(a-b)^2 + 4c^2} \right)\,.
	\end{align}
\end{subequations}
Note that by convention, $m_{\eta_1} \geq m_{\eta_2}$.~The DM candidate can either be the CP-even scalar $\eta_2$, the CP-odd pseudoscalar $A$ or the lightest Majorana fermion $\psi_2$ (as $m_{\psi_1} \geq m_{\psi_2}$ by convention). 

\section{Phenomenology} \label{sec:constraints}
After EWSB and rotation into the physical mass basis, the ScSM contains 12 free model parameters, namely
\begin{subequations}
	\begin{align}
		\textrm{5 masses (1 mixing angle)}: &\quad \{m_{\psi_1},\,m_{\psi_2},\,m_{\eta_1},\,m_{\eta_2},\,\,m_A,\,\theta\}\,, \\
		\textrm{6 couplings:} &\quad \{\lambda_\Phi,\,\lambda_{\varphi},\,\lambda_{H\Phi,1},\,\lambda_{H\Phi,2},\,\lambda_{H\varphi},\,\lambda_{\Phi\varphi}\}\,.
	\end{align}
\end{subequations}
The remaining parameters in eq.~\eqref{eqn:V11A} can be expressed as (see appendix~\ref{app:massbasis})
\begin{subequations}
\begin{align}
	m_\Phi^2 &= m_{\eta_1}^2 \cos^2 \theta + m_{\eta_2}^2 \sin^2 \theta - \frac{1}{2} \left(\lambda_{H\Phi,1} + \lambda_{H\Phi,2} + \lambda_{H\Phi,3} \right) \, v^2 \,, \\[1mm]
	m_\varphi^2 &= m_{\eta_1}^2 \sin^2 \theta + m_{\eta_2}^2 \cos^2 \theta - \frac{1}{2} \lambda_{H\varphi} v^2\,, \\[1mm]
	m_{\phi^+}^2 &= m_\Phi^2 + \frac{1}{2} \lambda_{H\Phi,1} v^2 \,, \\[1mm]
	\lambda_{H\Phi,3} &= \frac{1}{v^2} \left(m_{\eta_1}^2 \cos^2 \theta + m_{\eta_2}^2 \sin^2 \theta - m^2_A \right)\,,\\[1mm]
	\kappa &= \frac{1}{v} (m_{\eta_1}^2 - m_{\eta_2}^2) \sin \theta \cos\theta \,.
\end{align}
\end{subequations}
We also have free parameters within the complex Yukawa matrices, for which we use a Casas-Ibarra parametrisation~\cite{Casas:2001sr}.~For $N=2$ fermionic singlets, there are $N=2$ real angles from the complex orthogonal Casas-Ibarra matrix $R$, and $N-1=1$ Majorana phase in the PMNS matrix $U$ (as one of the neutrinos is massless); see appendix~\ref{app:yuk} for more details.~Thus, we can express the 6 complex Yukawa couplings in terms of the low-energy neutrino oscillation data (3 masses, 3 angles, and 2 phases), the 2 heavy Majorana masses parameters, and the 2 real angles of the $R$ matrix. These are considered as nuisance parameters in our study.

In the following subsections, we discuss various theoretical/observational constraints that are imposed on the allowed model parameter space. 

\subsection{Naturalness, perturbativity and $\mathbb{Z}_2$ symmetry breaking} \label{subsec:natpert}

\subsubsection{Perturbativity} \label{sec:pert}
From perturbativity arguments, we require all quartic couplings in the potential to satisfy
\begin{equation} \label{eq:Lpert}
	|\lambda_i| < 4 \pi\,.
\end{equation}
In addition, we also impose the co-positivity conditions discussed in appendix~\ref{app:stab}. As for the Yukawa couplings (derived parameters in our scan), we require
\begin{equation}
	|y^{\alpha}_{\psi_i}|^2 < 4 \pi\,,
\end{equation}
where $i=1$, $2$ and $\alpha =e,\,\mu,\,\tau$.

\subsubsection{Naturalness}
Using eq.~\eqref{eq:mixing} and expressing everything in terms of the physical scalar masses, the trilinear coupling $\kappa$ can be bounded as
\begin{equation}
	c \equiv \kappa v = (m^2_{\eta_1}-m^2_{\eta_2}) \sin\theta \cos\theta  \Longrightarrow | \kappa| \lesssim \frac{ \, m^2_{\eta_1}}{2v}\, \,,
\end{equation}
where we have used $m^2_{\eta_1} \gg m^2_{\eta_2}$ in the last step.~However, as the heavy neutral scalar masses are unknown, this upper bound is not very useful. 

We can make use of the fact that the trilinear coupling gives a correction to the Higgs boson mass at the one-loop level, where the new scalars run in the loop. Up to factors of 2 and logs, we estimate
\begin{equation}
	m^2_h \gtrsim\,\frac{|\kappa|^2}{16 \pi^2}\,.
\end{equation}
For the fine-tuning to be under control, i.e., $\delta m_h /m_h< \epsilon$, we demand that
\begin{equation} \label{eqn:nat}
	|\kappa| \lesssim4 \pi \epsilon \,m_h \simeq 1.5 \,{\rm TeV}\,,
\end{equation}
where we have assumed $\epsilon \simeq 1$. Similar considerations were made in the Zee model 
\cite{Herrero-Garcia:2017xdu}.~If the fine-tuning condition is relaxed (i.e., $\epsilon \gtrsim 1$), the upper limit on $\kappa$ can also be relaxed. 

\subsubsection{$\mathbb{Z}_2$ symmetry breaking at tree-level}
If $|\kappa|$ is much larger than the scalar masses, it can lead to a deeper minimum than the SM one, thereby breaking the $\mathbb{Z}_2$ symmetry.~Looking at different field directions, and using eq.~\eqref{eq:Lpert}, we get
\begin{equation}  \label{eq:Z2breaking}
	|\kappa| \lesssim \mathcal{O}(1)\, (-\mu_{H}^{2} + m_{\Phi}^{2}+\,m_{\varphi}^{2})\,.
\end{equation}
The above requirement, although more robust than the naturality one in eq.~\eqref{eqn:nat}, turns out to be weaker, particularly for $m_{\Phi}$ and $m_{\varphi}$ larger than the EW scale.~In order for the model to be valid above the EW scale (such that the $\mathbb{Z}_2$ symmetry is preserved), it is important to check that the RGE evolution does not only preserve $m_{\Phi}^{2}>0$ and $\,m_{\varphi}^{2}>0$, but also that $|\kappa|$ is not too large compared to the rest of the scalar masses.~The latter requirement, however, is expected to be easily satisfied, as $\kappa$ renormalises multiplicatively (see section \ref{sec:rge}).

\subsection{Neutrino masses} \label{subsec:numasses}

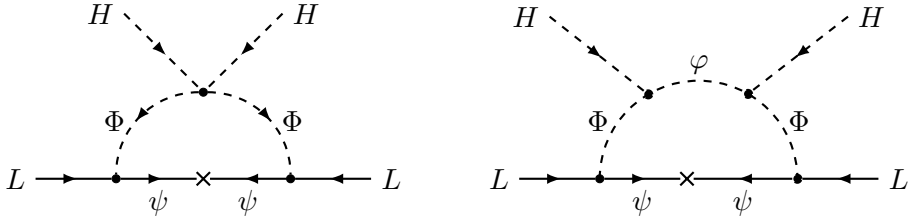
\begin{figure}[tb]
\centering
	\begin{tikzpicture}[node distance=1cm and 1cm]
     \coordinate[label=left:$L$] (nu1);
     \coordinate[vertex, right=of nu1] (v1);
     \coordinate[cross, right=of v1] (lfv);
     \coordinate[vertex, above=of lfv] (v3);
     \coordinate[above left=of v3,  label=left:$H$] (h1);
     \coordinate[above right=of v3,  label=right:$H$] (h2);
     \coordinate[vertex, right=of lfv] (v2);
     \coordinate[right=of v2, label=right:$L$] (nu2);
     \coordinate[above=of v1, xshift=0.25cm, yshift=-0.3cm, label=left:$\Phi$] (s1);
     \coordinate[above=of v2, xshift=-0.25cm, yshift=-0.3cm, label=right:$\Phi$] (s2);

     \draw[fermion] (nu1)--(v1);
     \draw[fermion] (v1) -- node[below]{$\psi$} ++ (lfv);
     \draw[fermion] (v2)-- node[below]{$\psi$} ++ (lfv);
     \draw[fermion] (nu2)--(v2);
     \draw[scalar] (h1) -- (v3);
     \draw[scalar] (h2) -- (v3);
     \draw[scalar] (v3) to[out=180,in=90] (v1);
     \draw[scalar] (v3) to[out=0,in=90] (v2);
   \end{tikzpicture}
 \qquad
  \centering
  \begin{tikzpicture}[node distance=1cm and 1cm]
   \coordinate[label=left:$L$] (nu1);
   \coordinate[vertex, right=of nu1] (v1);
   \coordinate[cross, right=of v1] (lfv);
   \draw[scalarnoarrow] (v1) arc[start angle=180, delta angle=-180, radius=1.3cm] node[pos=0.33,vertex] (v03) {} node[pos=0.667,vertex] (v07) {} node[vertex] (v2) {};
    \coordinate[right=of v2, label=right:$L$] (nu2);
    \coordinate[above=of nu1,xshift=0.4cm, yshift=1.15cm, label=left:$H$] (h1);
    \coordinate[above=of nu2,xshift=-0.4cm, yshift=1.15cm, label=right:$H$] (h2);
       \coordinate[above=of v1,xshift=0.25cm, yshift=-0.3cm,  label=left:$\Phi$] (s1);
     \coordinate[above=of v2, xshift=-0.25cm, yshift=-0.3cm,  label=right:$\Phi$] (s2);
      \coordinate[above=of lfv, xshift=-0.1cm, yshift=0.42cm,  label=right:$\varphi$] (s3);
      
    \draw[fermion] (nu1)--(v1);
      \draw[fermion] (v1) -- node[below]{$\psi$} ++ (lfv);
     \draw[fermion] (v2)-- node[below]{$\psi$} ++ (lfv);
    \draw[fermion] (nu2)--(v2);
    \draw[scalar] (h1)--(v03); 
    \draw[scalar] (h2)--(v07); 
  \end{tikzpicture}
%
\caption{Feynman diagrams for the neutrino masses at the one-loop level. The left panel shows the contribution in the Scotogenic Model (proportional to $\lambda_{H\Phi,3}$), whereas the right panel shows the new contribution in the ScotoSinglet Model (ScSM) (proportional to $\kappa^2$).}%
\label{fig:ScSM}
\end{figure}

Neutrino masses are generated at the one-loop level with the neutral fields running in the loop, see figure~\ref{fig:ScSM}.~From the Yukawa Lagrangian, eq.~\eqref{Lpsi}, and the scalar potential, eq.~\eqref{eqn:V11A}, we can see that the lepton number is violated by 2 units in the presence of $\mathbf{y_\Psi}$, $M_\Psi$ and either:
\begin{enumerate}
	\item The quartic coupling $\lambda_{H\Phi,3}\,$.~This contribution is same as in the ScM; see the left diagram in figure~\ref{fig:ScSM}.
	\item The square of the dimensionful trilinear coupling, $\kappa^2$. This is the new contribution in the ScSM; see the right diagram in figure~\ref{fig:ScSM}.
\end{enumerate}
These are the parameter combinations that enter in the expression for the neutrino masses, namely\footnote{We believe there are two typos in the expressions of ref.~\cite{Farzan:2009ji}: \emph{i}) a factor of $\pi$ is missing in the denominator; and \emph{ii}) an extra contribution to the scalar masses after EWSB is missing -- this is due to the $\lambda_{H \Phi,1}$ term in the potential ($\lambda_4$ in the notation of ref.~\cite{Farzan:2009ji}).}
\begin{align} \label{neutrinomassmatrix}
	(M_\nu)_{\alpha\beta} = \sum_{k=1}^2 \frac{y_{k \alpha} \, m_{\psi_k} \, y_{k \beta} }{32\pi^2} \left[ \cos^2\theta \, F_k(m_{\eta_1}) + \sin^2\theta \, F_k(m_{\eta_2}) - F_k(m_A) \right] \,,
\end{align}
where the loop function is
\begin{align} \label{eq:loop}
	F_k(m_x) = \frac{m_x^2}{m_x^2 - m_{\psi_k}^2} \log \left(\frac{m_x^2}{m_{\psi_k}^2}\right) \,.
\end{align}
It is instructive to expand in the limit of small $ \lambda_{H \Phi,3} $ and $ \kappa $. For $ a>b $ and small $\kappa$, the mixing is $ \theta \approx 0 $, thus $ \eta_1 \, (\eta_2)$  is mostly doublet (singlet).~In this case, the mass splitting between $ \eta_1 $ and $ A $ reads $ m_{\eta_1}^2-m_A^2\approx \lambda_{H \Phi,3} v^2 $. For $ m_{\eta_1}>m_{\eta_2}\gg m_{\psi_k} $, we find that
\begin{align}
	(M_\nu)_{\alpha\beta} \approx \sum_{k=1}^2 \frac{y_{k \alpha} \, m_{\psi_k} \, y_{k \beta} }{32\pi^2} \left[ \frac{\lambda_{H \Phi,3} \,v^2}{m_{\eta_1}^2}  - \frac{\kappa^2v^2}{(m_{\eta_1^2}-m_{\eta_2}^2)^2}\log\left(\frac{m_{\eta_1}^2}{m_{\eta_2}^2}\right) \right] \,.
\end{align}
For $ m_{\eta_1}\gg m_{\psi_k}\gg m_{\eta_2} $, we get
\begin{align}
	(M_\nu)_{\alpha\beta} \approx \sum_{k=1}^2 \frac{y_{k \alpha} \, m_{\psi_k} \, y_{k \beta} }{32\pi^2} \left[ \frac{\lambda_{H \Phi,3}\, v^2}{m_{\eta_1}^2}  - \frac{\kappa^2v^2}{m_{\eta_1}^4}\log\left(\frac{m_{\eta_1}^2}{m_{\psi_k}^2}\right) \right] \,.
\end{align}
Alternatively, for $ b>a $, $ \theta \approx \pi/2 $ and thus $ \eta_1\,(\eta_2)$ is mostly singlet (doublet). The mass splitting between $ \eta_2 $ and $ A $ is $ m_{\eta_2}^2-m_A^2\approx \lambda_{H \Phi,3} v^2 $. For $ m_{\eta_1}>m_{\eta_2}\gg m_{\psi_k} $, we get
\begin{align}
	(M_\nu)_{\alpha\beta} \approx \sum_{k=1}^2 \frac{y_{k \alpha} \, m_{\psi_k} \, y_{k \beta} }{32\pi^2} \left[ \frac{\lambda_{H \Phi,3} \,v^2}{m_{\eta_2}^2}  + \frac{\kappa^2v^2}{(m_{\eta_1^2}-m_{\eta_2}^2)^2}\log\left(\frac{m_{\eta_1}^2}{m_{\eta_2}^2}\right) \right] \,,
\end{align}
whereas for $ m_{\eta_1}\gg m_{\psi_k}\gg m_{\eta_2} $, we get
\begin{align}
	(M_\nu)_{\alpha\beta} \approx \sum_{k=1}^2 \frac{y_{k \alpha} \, m_{\psi_k} \, y_{k \beta} }{32\pi^2} \left\{ \frac{\lambda_{H \Phi,3} v^2}{m_{\psi_k}^2}\left[\log\left(\frac{m_{\psi_k}^2}{m_{\eta_2}^2}\right)-1\right]  + \frac{\kappa^2v^2}{m_{\eta_1}^4}\log\left(\frac{m_{\eta_1}^2}{m_{\psi_k}^2}\right) \right\}\,.
\end{align}
Note how the Scotogenic-like contribution (proportional to $\lambda_{H \Phi,3}\, v^2$) is always suppressed by the fermion or doublet mass, depending on which one is the heaviest state in the spectrum.

\subsection{Integrating-out the heavy scalar singlet}
If the singlet $\varphi$ is the heaviest particle in the spectrum, e.g., $m_{\varphi}^2 \gg m_\Phi^2$  ($b \gg a$), it can be integrated out (see ref.~\cite{Casas:2017jjg} for a similar study, and ref.~\cite{Bilenky:1993bt} for the example of a charged scalar singlet). For scalar DM and $\theta \simeq \pi/2$, the DM candidate $\eta_2$ is mainly doublet and this is a good approximation.~Assuming that the weak and mass eigenstates are similar, i.e., for small mixing (small $\theta$), after integrating out the singlet at tree-level before EWSB, we obtain the following expression for the scalar potential at dimension-6:
\begin{align}\label{eq:Vb}
	V_{\rm eff} (H, \Phi) &= \left(\lambda_{H\Phi,2} -\frac{|\kappa|^2}{m_{\varphi}^2}\right) |H^\dagger \Phi|^2 +\frac{1}{2} \left[\left(\lambda_{H\Phi,3} -\frac{(\kappa^\dagger)^2}{m_{\varphi}^2}\right) (H^\dagger \Phi)^2 + {\rm H.c.} \right] \nonumber \\
	&\hspace{5mm} -\frac{\lambda_{H\varphi}\,|\kappa|^2}{m_{\varphi}^4} |H^\dagger \Phi|^2 H^\dagger H -\frac{\lambda_{\Phi\varphi}\,|\kappa|^2}{m_{\varphi}^4} |H^\dagger \Phi|^2 \Phi^\dagger \Phi \nonumber \\
	&\hspace{5mm} -\frac{1}{2} \left(\frac{\lambda_{H\varphi}\,(\kappa^\dagger)^2}{m_{\varphi}^4} (H^\dagger \Phi)^2 H^\dagger H + {\rm H.c.} \right) \nonumber \\
	&\hspace{5mm} - \frac{1}{2} \left(\frac{\lambda_{\Phi\varphi}\,(\kappa^\dagger)^2}{m_{\varphi}^4} (H^\dagger \Phi)^2 \Phi^\dagger \Phi + {\rm H.c.} \right)\,.
\end{align}
After EWSB, the effective coupling $\dfrac{1}{2}\, [\lambda_{H\Phi,3}^{\rm eff}\, (H^\dagger \Phi)^2 + {\rm H.c.}]$ that appears in the neutrino masses is 
\begin{equation} \label{eq:L3eff}
	\lambda_{H\Phi,3}^{\rm eff} \equiv \lambda_{H\Phi,3} -\frac{(\kappa^\dagger)^2}{m_{\varphi}^2} - \lambda_{H\varphi} \frac{(\kappa^\dagger)^2}{m_{\varphi}^4} \frac{v^2}{2}\,.
\end{equation}
This agrees with our expectation from considerations of lepton number violation.~We see that neutrino masses can be suppressed either by small $\lambda_{H\Phi,3}$ and $\kappa$, or by cancellations among these terms.~Notice that the last term in eq.~\eqref{eq:L3eff} gives a contribution to neutrino masses via a dimension-7 Weinberg-like operator, which is expected to be suppressed compared to the usual dimension-5 one (the rest of the terms). In our numerical scan, we check that the combination in eq.~\eqref{eq:L3eff} is indeed fixed (with the scale set by the neutrino masses).

Let us elaborate a bit more on the threshold corrections to $\lambda_{H\Phi,2}$ and $\lambda_{H\Phi,3}$ at the scale of the singlet $\varphi$.~Interestingly, the threshold effects increase their values for high energies. The stability conditions, however, must to be applied differently above and below $m_\varphi$.~This effect has been used to keep the Higgs potential stable up to high energy scales in ref.~\cite{EliasMiro:2012ay}. Notice also that the signs of $\lambda_{H\Phi,3}$ and $\kappa$ are preserved under the RGE flow due to their multiplicative renormalization.~The behaviour in eq.~\eqref{eq:L3eff} can also be understood from the  scalar mass matrix, see eqs.~\eqref{eqn:nondia_matrix} and \eqref{eq:scmass}.~In the symmetric Higgs phase (scales above the EW scale) and for $m_\varphi^2 \gg m_\Phi^2$, the neutral scalar masses are see-saw like, i.e., from eqs.~\eqref{eqs:massesab}, we have
\begin{equation}
	m_{\eta_1}^2 \simeq m^2_\varphi + \frac{\kappa^2 \, v^2}{m_\varphi^2}\,, \quad m_{\eta_2}^2 \simeq m^2_\Phi - \frac{\kappa^2 \, v^2}{m_\varphi^2}\,.
\end{equation}
We see that the lightest state (by convention $\eta_2$) is mainly doublet ($\Phi$), while the heaviest state ($\eta_1$) is mainly singlet ($\varphi$), corresponding to a mixing angle of $ \theta \approx \pi/2 $. This repulsion of mass eigenvalues is the same effect that we see in the quartic couplings $\lambda_{H\Phi,2}$ and $\lambda_{H\Phi,3}$, see eq.~\eqref{eq:Vb}.

\subsection{Electroweak precision tests} \label{subsec:ewpt}
The new particles in the ScSM contribute to the $W^\pm$ and $Z$ boson self-energies. These contributions are parametrised by the oblique $S$, $T$ and $U$ parameters \cite{PhysRevD.46.381,PhysRevLett.65.964}. The strongest constraint comes from the $T$ parameter, which bounds the mass splitting of the neutral and charged scalars. It is given by \cite{Haber:2010bw}
\begin{align}
	T  &= \dfrac{1}{16\pi^2 \alpha_{\rm em} v^2} \biggl[\cos^2 \theta \, \mathcal{F}(m_{\phi^+}^2,\,m_{\eta_1}^2) + \sin^2\theta \, \mathcal{F}(m_{\phi^+}^2,\,m_{\eta_2}^2) + \mathcal{F}(m_{\phi^+}^2,\,m_A^2) \nonumber \\
	&\hspace{5mm} - \cos^2\theta \, \mathcal{F}(m_{\eta_1}^2,\,m_{A}^2) -\sin^2\theta \, \mathcal{F}(m_{\eta_2}^2,\,m_{A}^2) \biggr] \,,
\end{align}
where the loop function (symmetric in $x$ and $y$) is 
\begin{equation}
	\mathcal{F}(x^2,\,y^2) = \dfrac{x^2+ y^2}{2} - \dfrac{x^2 y^2}{x^2-y^2}\ln\left(\dfrac{x^2}{y^2}\right)\,.
\end{equation}
The $S$ parameter is given by
\begin{equation} \label{spar}
	S = \frac{1}{\pi m_Z^2} \biggl[\cos^2\theta \, \mathcal{B}_{22}(m_Z^2,\,m_{\eta_1}^2,\,m_{A}^2) + \sin^2\theta \, \mathcal{B}_{22}(m_Z^2,\,m_{\eta_2}^2,\,m_A^2) -\mathcal{B}_{22}(m_Z^2,\,m^2_{\phi^+},\,m^2_{\phi^+})\biggr] \,.
\end{equation}
Similarly, the combination $S+U$ reads
\begin{align} \label{s+upar}
	S+ U &= \frac{1}{\pi m_W^2} \biggl[\cos^2\theta \, \mathcal{B}_{22}(m_W^2,\,m^2_{\phi^+},\,m_{\eta_1}^2) + \sin^2\theta \, \mathcal{B}_{22}(m_W^2,\,m^2_{\phi^+},\,m_{\eta_2}^2) \nonumber \\
	&\hspace{5mm} +  \mathcal{B}_{22}(m_W^2,\,m^2_{\phi^+},\,m_A^2) -2\,\mathcal{B}_{22}(m_W^2,\,m^2_{\phi^{+}},\,m^2_{\phi^{+}}) \biggr]\,,
\end{align}
where
\begin{equation}
	\mathcal{B}_{22}(q^2,\,m_1^2,\,m_2^2) = B_{22}(q^2,\,m_1^2,\,m_2^2)-B_{22}(0,\,m_1^2,\,m_2^2)\,.\label{b22} 
\end{equation}
The Passarino--Veltman function $B_{22}$~\cite{Passarino:1978jh} (symmetric in the last two arguments) is
\begin{align} 
	\label{B}
	B_{22}(q^2,\,m_1^2,\,m_2^2) &= \frac{1}{4}(\Delta+1)(m_1^2+m_2^2-\frac{1}{3}q^2)-\frac{1}{2}\int^1_0 X \ln(X-i\epsilon) \, {\rm d}x \,,
\end{align}
where 
\begin{equation}
	X \equiv m_1^2 x + m_2^2(1-x) -q^2x(1-x)\,, \quad
	\Delta \equiv \frac{2}{4-d}+\ln 4\pi-\gamma_E\,,
\end{equation}
in $d$ space-time dimensions and $\gamma_E \simeq 0.577$ is the Euler--Mascheroni constant.~We use the compact analytic expressions from appendix~B of ref.~\cite{Herrero-Garcia:2017xdu}.
Our expressions agree with the ones for the Inert Doublet Model (IDM) \cite{Barbieri:2006dq} in the appropriate limits.

The oblique parameters are constrained from the global electroweak fit \cite{Haller:2018nnx}, assuming SM reference values of $m_h^{\textnormal{ref}} = 125$\,GeV and $m_t^{\textnormal{ref}} = 172.5$\,GeV. The most recent fit gives
\begin{equation}\label{eqn:STU_pars}
	S = 0.04 \pm 0.11, \quad T = 0.09 \pm 0.14, \quad U = -0.02 \pm 0.11,
\end{equation}
along with the following correlation matrix:
\begin{equation}\label{eqn:corr_mat}
	\rho_{ij} = 
	\begin{pmatrix}
		1 & 0.92 & -0.68 \\
		0.92 & 1 & -0.87 \\
		-0.68 & -0.87 & 1
	\end{pmatrix}.
\end{equation}

\subsection{Higgs decay into di-photons}\label{subsec:diphoton}
The coupling between the SM Higgs $h$ and charged scalar $\phi^+$ in eq.~\eqref{eqn:V11A} modifies the decay rate of the $h \rightarrow \gamma \gamma$ process. The ratio with respect to the SM value is  \cite{Ellis:1975ap,Shifman:1979eb,Carena:2012xa}
\begin{equation}
	\mathcal{R}_{\gamma \gamma} \equiv \frac{\Gamma(h \to \gamma \gamma)_{\rm ScSM}}{\Gamma(h \to \gamma \gamma)_{\rm SM}} = \left |1 + \frac{\lambda_{H \Phi,1} \, v^2}{2  m_{\cal \phi^+}^2}\frac{A_0(\tau_{\phi^+})}{A_1(\tau_W) + \dfrac{4}{3} A_{1/2}(\tau_t)} \right |^2\,. \label{eq:ggratio}
\end{equation}
Here the loop functions $A_i (\tau_j \equiv 4 m_j^2/m_h^2)$ read
\begin{subequations}
\begin{align}
	A_0(x) &= -x+x^2 \, f \left (\frac{1}{x}\right )\,, \\
	A_{1/2}(x) &= 2x+ 2x (1- x) \, f \left (\frac{1}{x}\right )\,, \\
	A_{1}(x) &= -2-3x-3x (2-x) \, f \left (\frac{1}{x}\right )\,,
\end{align}
\end{subequations}
where $j = \phi^+$, $t$ and $W$, and $f=\arcsin^2(\sqrt x )$ for $m_h<2 m_i$.

\subsection{Lepton flavour violation} \label{subsec:LFV}
We use the usual expressions for the lepton flavour violation (LFV) processes (including $\mu - e$ conversion rates in various elements) that are applicable for the ScM~\cite{Toma:2013zsa,Hagedorn:2018spx}. For instance, radiative decays are given by 
\begin{equation}
	\mathcal{BR}(\ell_\alpha \rightarrow \ell_\beta \gamma) = 
\frac{3 \, \alpha_{\rm em} }{64 \pi \, G^2_F \, m^4_{\phi^+}} \left|\sum_i\,y^*_{i \beta}\,y_{i \alpha}\, F\left(\frac{m^2_{\psi_i}}{m^2_{\phi^+}}\right) \right|^2\, \mathcal{BR}(\ell_\alpha \rightarrow \ell_\beta    \nu_\alpha \overline{\nu_\beta})\,,
\end{equation}
where 
\begin{equation}
F(x) = \frac{1- 6x+3x^2 +2x^3 - 6x^2 \log x}{6(1-x)^4}\,.
\end{equation}
The various LFV processes that we include in our study are summarised in table~\ref{tab:BRs_LFV}. 

\begin{table} [t!]
	\begin{center}
	  	\begin{tabular}{c|cc}
    		\hline \hline
    		\textbf{LFV process} & \textbf{Upper limit (90\%\,CL)} & \textbf{Ref.} \\ \hline
    		$\mu \rightarrow e \gamma$ & $4.2 \times 10^{-13}$ & MEG \cite{TheMEG:2016wtm}\\
    		$\tau \rightarrow \mu \gamma$ & $4.4 \times 10^{-8}$ & PDG \cite{Tanabashi:2018oca}  \\
    		$\tau \rightarrow e \gamma$ & $3.3 \times 10^{-8}$ & PDG \cite{Tanabashi:2018oca}  \\ \hline
    		$\mu \rightarrow 3 e$ & $1.0 \times 10^{-12}$ & PDG \cite{Tanabashi:2018oca}  \\
    		$\tau \rightarrow 3 \mu$ & $2.1 \times 10^{-8}$ & PDG \cite{Tanabashi:2018oca}  \\
    		$\tau \rightarrow 3 e$ & $2.7 \times 10^{-8}$ & PDG \cite{Tanabashi:2018oca}  \\ \hline 
    		$\mu \rightarrow e ~ (\mathsf{Au})$ & $7.0 \times 10^{-13}$ & PDG \cite{Tanabashi:2018oca}  \\ 
    		$\mu \rightarrow e ~ (\mathsf{Ti})$ & $4.3 \times 10^{-12}$ & PDG \cite{Tanabashi:2018oca} \\ \hline \hline
    	\end{tabular}
		\caption{Upper limits on the branching ratios/conversion rates of various lepton flavour violation (LFV) processes at 90\%\,CL. Here $\mu \rightarrow e$ refers to conversion rate in gold ($\mathsf{Au}$) and titanium ($\mathsf{Ti}$).}
		\label{tab:BRs_LFV}
	\end{center}	
\end{table}

\subsection{Relic abundance} \label{subsec:relic}
The ScSM permits both scalar and fermionic DM candidates, but we focus on the case of scalar DM for two reasons: 
\begin{enumerate}
	\item The fermion DM case is very similar to the usual ScM where strong constraints from LFV exist on the Yukawa couplings. It requires special textures, and/or coannihilations and/or fermion triplets instead of singlets to saturate the observed DM abundance. In our scan, we indeed find the need for coannihilations (see section~\ref{sec:results}).
	\item It is interesting to study the rich phenomenology of CP-even scalar DM $(\eta_2)$ as an admixture of singlet (pure singlet case is $\theta \simeq 0$) and doublet (pure doublet case is $\theta \simeq \pi/2$) components, see eq.~\eqref{eqn:mixingSC}. The doublet case (both CP-even and CP-odd) has the phenomenology of the ScM, while the singlet has some extra terms from the scalar potential with respect to just a pure singlet scalar.~As the CP-odd scalar $A$ can be a DM candidate, there are a few possible mass hierarchies:
	\begin{itemize}
		\item DM candidate = $\eta_2$ with $m_{\eta_2}< m_{\eta_1} < m_A$ or $m_{\eta_2} < m_A < m_{\eta_1}$;
		\item DM candidate = $A$ with $m_A < m_{\eta_2} < m_{\eta_1}$.
	\end{itemize}
\end{enumerate}

In general, several (co-)annihilation channels are possible in the ScSM. All of these are included in \textsf{micrOMEGAs v5.2.0} \cite{Belanger:2018mqt} which we use to compute the DM relic abundance. The DM abundance is required to be equal to or smaller than the \emph{Planck} (2018) measured abundance \cite{Aghanim:2018eyx}: 
\begin{equation}\label{eqn:relic}
	\Omega_{\rm DM} h^2 = 0.120 \pm 0.001.
\end{equation}
Thus, the \emph{Planck} measurement provides an upper limit on the DM relic abundance. 

\subsection{Direct detection} \label{subsec:DD}
Direct detection (DD) typically imposes strong constraints on scalar DM candidates with a non-zero hypercharge due to the presence of $t$-channel $Z$/$h$-mediated diagrams~\cite{Hambye:2009pw}. The gauge interactions stem from the kinetic term for the doublet, namely
\begin{align}  
	\mathscr{L} \supset& \, (D_\mu \Phi)^\dagger (D^\mu \Phi) \nonumber \\
	\supset & - \frac{g}{2 \cos \theta_W} Z_\mu \left( \phi_R\,\partial^\mu A - A\,\partial^\mu \phi_R \right) \nonumber \\
	= & - \frac{g}{2 \cos \theta_W} Z_\mu \Big[  \cos\theta (\eta_1\,\partial^\mu A - A\, \partial^\mu \eta_1) - \sin\theta (\eta_2 \, \partial^\mu A - A \, \partial^\mu \eta_2) \Big]\,. \label{eq:Z}
\end{align}
Notice that the $Z$-mediated interactions are inelastic, and always involve a CP-even scalar and pseudoscalar $A$.~A small enough mixing can suppress DD limits via the last term in eq.~\eqref{eq:Z}, as in this case, $\eta_2$ is mainly singlet and does not directly couple to the $Z$-boson.~In addition, a large enough mass splitting ($\gtrsim$ MeV) can kinematically forbid the scatterings between $\eta_2$ and $A$. This implies that $\lambda_{H\Phi,3} \gtrsim 10^{-6}$. Two cases are possible: 
\begin{enumerate}
	\item For $b \gg a$ (and $b \gg c$), $\eta_2$ is mainly doublet (with a correction proportional to $\kappa$) and the mass splitting with $A$ is given by $\lambda_{H \Phi,3}$ (which can be made naturally smaller than in the ScM);
	\item For $b \ll a$ (and $a \gg c$), $\eta_2$ is mainly singlet and $m_A$ is given in terms of other parameters, so there is no reason for the mass splitting to be small, and thus inelastic scatterings are expected to be forbidden. 
\end{enumerate}

Interactions mediated by the SM Higgs $h$ lead to the usual elastic (and also inelastic) spin-independent (SI) scattering; the resulting limits are also quite severe \cite{Athron:2018ipf,Athron:2018hpc}, but they can be suppressed by small scalar couplings unlike in the case of $Z$-mediated interaction \cite{Escudero:2016gzx}. For non-zero mixing, the dimensionless $h$-$\eta_2$-$\eta_2$ coupling is given by \cite{Cohen:2011ec}
\begin{equation}\label{eqn:DM-coupling}
	\lambda_{\mathrm{eff}} \equiv \frac{1}{v} \left[\lambda_{H\varphi} v \cos^2 \theta - 2 \kappa \sin\theta\cos\theta + \lambda_{123} \, v \sin^2\theta \right] \,,
\end{equation}
where $ \lambda_{123} \equiv \lambda_{H \Phi,1} + \lambda_{H\Phi,2} + \lambda_{H \Phi,3}$.~As $\lambda_{\mathrm{eff}}$ decreases, the $h$-mediated direct detection cross section also decreases.~This coupling will vanish (i.e., no overall $h$-$\eta_2$-$\eta_2$ coupling) when the mixing angle satisfies the following relation:
\begin{align} \label{eqn:cancel}
	\sin^2\theta = \frac{(m_\Phi^2 - m_\varphi^2)+\lambda_{H\varphi} v^2 - \sqrt{(m_\Phi^2 - m_\varphi^2)^2 + \lambda_{123}\lambda_{H\varphi}v^4}}{2 \, (m_\Phi^2 - m_\varphi^2)-\lambda_{123} v^2 + \lambda_{H\varphi} v^2} \,,
\end{align}
We confirm that when the coupling $ \lambda_{H\varphi} $ ($ \lambda_{123}$) vanishes, the Higgs-DM coupling is zero only for pure singlet (doublet) DM. In the event that the couplings are exactly equal and dominate over the bare masses, $ \lambda_{123} = \lambda_{H\varphi} \gg |m_\Phi^2-m_\varphi^2|/v^2 $, the effective Higgs coupling $\lambda_{\mathrm{eff}}$ vanishes for maximal mixing $ \theta = \pi/4 $.

Using \textsf{micrOMEGAs}, we compute the effective SI DM-proton scattering cross section, assuming that the local DM energy density scales proportional to the global one:
\begin{equation}\label{eqn:effSI}
	\sigma_{\mathrm{eff}} \equiv \sigma_{\mathrm{SI}}^p \cdot f_{\mathrm{rel}}\,, \quad f_{\mathrm{rel}} \equiv \Omega_X/\Omega_{\mathrm{DM}}\,,
\end{equation}
where $\Omega_X$ is the $X$ DM abundance ($X = \eta_2$, $A$ or $\psi_2$) and $\Omega_{\mathrm{DM}}$ is the \emph{Planck} measured abundance in eq.~\eqref{eqn:relic}.~We then recast the observed exclusion limit from XENON1T \cite{Aprile:2018dbl} within \textsf{micrOMEGAs} \cite{Belanger:2020gnr} for a fixed DM mass and compare it against the effective SI cross section in eq.~\eqref{eqn:effSI}.~There are a number of planned xenon-based experiments that will increase the sensitivity significantly, e.g., XENONnT~\cite{Aprile:2014zvw}, PandaX~\cite{Cui:2017nnn}, LZ~\cite{Akerib:2015cja} and DARWIN~\cite{Aalbers:2016jon}. In our plots, we will illustrate the expected reach of LZ only.

\subsection{$\mathbb{Z}_2$ symmetry at high energies} \label{sec:rge}
The ScSM also has interesting features in light of the model viability up to high-energy scales \cite{Merle:2015gea,Merle:2015ica,Lindner:2016kqk,Escribano:2020iqq}.~The trilinear coupling $ \kappa $, due to the presence of the scalar singlet, gives positive contributions to the RGE  evolution of the new scalar mass-squared parameters if it is real, thereby helping prevent the breaking of the model symmetries even in the absence of finite temperature effects.~The compatibility with the evolution of the bare Higgs mass (and thus of EWSB) requires a separate study, which we leave for a future work. In principle, one could incorporate the evolution of Renormalisation Group Equations (RGEs) of model parameters into our numerical scan to verify that the $ \mathbb{Z}_2 $ symmetry remains unbroken at high-energy scales, but the practical implementation remains computationally difficult. Thus, we do not include RGE effects in our numerical analysis. 

\section{Numerical analysis}\label{sec:scandet}
To efficiently sample the allowed parameter space of the ScSM, we use the Importance Nested Sampling algorithm implemented in \textsf{MultiNest v3.10.0} \cite{Feroz:2008xx} with 25,000 live points (\textsf{nlive}) and a stopping tolerance (\textsf{tol}) of 10$^{-3}$.\footnote{As we will see in section~\ref{sec:results}, our fixed log-likelihood contours are mostly flat in the model parameter planes.~For this reason, we run \textsf{MultiNest} with stringent settings to efficiently sample the $1
\sigma$ and $2\sigma$ CL regions.}~The composite log-likelihood used is
\begin{align}\label{eqn:tot_like}
	\ln \mathcal{L}_{\rm{total}} (\bm{\theta}) &= \ln \mathcal{L}_\kappa (\bm{\theta}) + \ln \mathcal{L}_{\rm{EWPT}} (\bm{\theta}) 
	+ \ln \mathcal{L}_{\mathcal{R}_{\gamma \gamma}} (\bm{\theta}) \nonumber \\
	&\hspace{5mm} + \ln \mathcal{L}_{\rm{LFV}} (\bm{\theta}) + \ln \mathcal{L}_{\Omega h^2} (\bm{\theta}) + \ln \mathcal{L}_{\rm{DD}} (\bm{\theta}) \,,
\end{align}
where $\bm{\theta}$ are the free parameters of the ScSM. The individual log-likelihood contributions are described below:
\begin{enumerate}
	\item $\ln \mathcal{L}_\kappa (\bm{\theta})$:~log-likelihood for the trilinear coupling $\kappa$. It is Gaussian in nature, centered at 0 TeV with a standard deviation of 1.5 TeV, see eq.~\eqref{eqn:nat}.
	\item $\ln \mathcal{L}_{\rm{EWPT}} (\bm{\theta})$:~log-likelihood for the electroweak precision tests (EWPT) (see subsection~\ref{subsec:ewpt}). It is given by \cite{Profumo:2014opa}
	\begin{equation}
	    \ln \mathcal{L}_{\textnormal{EWPO}} (\bm{\theta}) = -\frac{1}{2} \sum_{i,\,j} (\Delta \mathcal{O}_i - \overline{\Delta \mathcal{O}}_i) \left(\Sigma^2 \right)_{ij}^{-1} (\Delta \mathcal{O}_j - \overline{\Delta \mathcal{O}}_j),
	\end{equation}
	where $\overline{\Delta \mathcal{O}}_i$ are the central values for the shifts in eq.~\eqref{eqn:STU_pars}, $\Sigma_{ij}^2 \equiv \sigma_i \rho_{ij} \sigma_j$ is the covariance matrix, $\rho_{ij}$ is the correlation matrix in eq.~\eqref{eqn:corr_mat} and $\sigma_i$ are the associated errors in eq.~\eqref{eqn:STU_pars}.
	\item $\ln \mathcal{L}_{\mathcal{R}_{\gamma \gamma}} (\bm{\theta})$:~log-likelihood for $R_{\gamma \gamma} \equiv \Gamma(h \to \gamma \gamma)_{\rm ScSM}/\Gamma(h \to \gamma \gamma)_{\rm SM}$ (see subsection~\ref{subsec:diphoton}). It is a Gaussian likelihood function, centered at the PDG measured value of 1.1 with a standard deviation of 0.1 \cite{Tanabashi:2018oca}.~The SM expectation is $\mathcal{R}_{\gamma \gamma} = 1.0$.
	\item $\ln \mathcal{L}_{\rm{LFV}} (\bm{\theta})$:~log-likelihood for the LFV processes (see subsection~\ref{subsec:LFV}). It is given by
	\begin{align}
		\ln \mathcal{L}_{\rm LFV} (\bm{\theta}) &=  \ln \mathcal{L}_{\mu\,\rightarrow\,e \gamma} (\bm{\theta}) + \ln \mathcal{L}_{\tau\,\rightarrow\,\mu \gamma}(\bm{\theta}) + \ln \mathcal{L}_{\tau\,\rightarrow \, e \gamma} (\bm{\theta}) \nonumber \\
		&\hspace{5mm} + \ln \mathcal{L}_{\mu\,\rightarrow\,3e} (\bm{\theta}) + \ln \mathcal{L}_{\tau\,\rightarrow\,3\mu}(\bm{\theta}) + \ln \mathcal{L}_{\tau\,\rightarrow \, 3e} (\bm{\theta}) \nonumber \\
		&\hspace{5mm} + \ln \mathcal{L}_{\mu \,\rightarrow \,e \,\, (\mathsf{Au})} (\bm{\theta}) + \ln \mathcal{L}_{\mu \, \rightarrow \, e \,\, (\mathsf{Ti})} (\bm{\theta}) \,.
	\end{align}
	Each of the individual likelihood functions are Gaussian, and centered at a branching ratio/conversion rate of 0 with a standard deviation equal to the respective upper limit shown in column 2 of table~\ref{tab:BRs_LFV}.
	\item $\ln \mathcal{L}_{\Omega h^2} (\bm{\theta})$:~log-likelihood for the DM relic density $\Omega h^2$ (see subsection~\ref{subsec:relic}). It is a one-sided Gaussian, i.e., a flat likelihood on $\Omega_X \leq \Omega_{\mathrm{DM}}$ and Gaussian for $\Omega_X > \Omega_{\mathrm{DM}}$. The \emph{Planck} measured uncertainty is also combined in quadrature with a 5\% theoretical uncertainty (stemming from our assumed uncertainty on the relic density calculation in \textsf{micrOMEGAs}).
	\item $\ln \mathcal{L}_{\rm{DD}} (\bm{\theta})$:~log-likelihood for the XENON1T experiment (see subsection~\ref{subsec:DD}).~It is a simple step-function-like likelihood, i.e., parameter points are allowed (rejected) if the effective SI cross section in eq.~\eqref{eqn:effSI} is below (above) the official XENON1T exclusion limit \cite{Aprile:2018dbl} for a given DM mass.
\end{enumerate}

The ranges and priors for the 21 (12 free + 9 nuisance) model parameters in normal ordering (NO) and inverted ordering (NO) are summarised in table~\ref{tab:param}.\footnote{We keep the CP-even $(m_{\eta_{1,2}})$ and CP-odd $(m_A)$ scalar masses $\gtrsim 100$\,GeV to avoid constraints from $h/Z$ invisible decays and collider limits.}~Due to the presence of coannihilations, we find it efficient to scan over $ \delta_1 $ and $ \delta_A $ (instead of $ m_{\eta_1} $ and $ m_A $) where
\begin{align}
	\delta_1 \equiv m_{\eta_1} - m_{\eta_2} \,, \quad \delta_A \equiv m_A - m_{\eta_2} \,.
\end{align}
By convention, $\delta_1 \geq 0$ as $m_{\eta_1} \geq m_{\eta_2}$. Meanwhile, $ \delta_A > 0$ for $\eta_2$ as DM, $\delta_A < 0$ for $A$ as DM, and either for $ \psi_2 $ as DM. 

\begin{table}[t]
	\begin{center}
  		\begin{tabular}{l|rc}
		    \hline \hline
		    & \hspace{0.37cm}\textbf{Ranges} & \textbf{Priors} \\ \hline
		    \textbf{Model parameters} & & \\
		    $\{m_{\psi_1},\,m_{\psi_2}\}$~(GeV) & \hspace{0.37cm}$[10,\,10^{6}]$ & Log \\
		    $\,m_{\eta_2}$\,(GeV) & \hspace{0.37cm}$[100,10^{4}]$ & Log \\
		    $\delta_1 \equiv m_{\eta_1} - m_{\eta_2}$\,(GeV) & ~~~$[10^{-3},\,10^4]$ & Log \\
		    $\delta_A \equiv m_A - m_{\eta_2}$\,(GeV) & ~~~$[-10^{4},\,-10^{-3}]\cup[10^{-3}, 10^4]$ & Log $|$value$|$ \\
		    $\theta$\,(rad.) & \hspace{0.37cm}$[0,\,\pi]$ & Flat \\
		    $\{\lambda_\Phi,\,\lambda_{\varphi}\}$ & \hspace{0.37cm}$[10^{-3},\,4\pi]$ & Log  \\
		    $\{\,\lambda_{H\Phi,1},\,\lambda_{H\Phi,2},\,\lambda_{H\varphi},\,\lambda_{\Phi\varphi}\}$ & $[-4\pi,\,-10^{-3}]\cup[10^{-3}, 4\pi]$ & Log $|$value$|$ \\[1mm] \hline
		    \textbf{Nuisance parameters} & & \\
		    $\sin^2 \theta_{12}$ & \hspace{0.33cm}$[0.275,\,0.350]$ & Flat \\[2mm]
		    \multirow{2}{*}{$\sin^2 \theta_{13}$} & $[0.02044,~0.02435]$~(NO) \\ 
		    & $[0.02064,~0.02457]$~(IO) & Flat \\[2mm]
		    \multirow{2}{*}{$\sin^2 \theta_{23}$} & $[0.433,~0.609]$~(NO) \\ 
		    & $[0.436,~0.610]$~(IO) & Flat \\[2mm]
		    $\dfrac{\Delta m_{21}^2}{10^{-5}\,\mathrm{eV}^2}$ & \hspace{0.35cm}$\left[6.79,~8.01\right]$ & Flat \\[2mm]
		    \multirow{2}{*}{$\dfrac{\Delta m_{3l}^2}{10^{-3}\,\mathrm{eV}^2}$} & $\left[2.436, 2.618\right]$~(NO) \\ 
		    & $\left[-2.601, -2.419\right]$~(IO) & Flat \\[2mm]
			\multirow{2}{*}{$\delta_{\rm CP}~(^{\circ})$} & $[144,~357]$~(NO) \\ 
		    & $[205,~348]$~(IO) & Flat \\[2mm]
		    $\alpha$\,(rad.) & \hspace{0.33cm}$[0,\,2\pi]$ & Flat \\
		    $\{\zeta_1,\,\zeta_2\}$ & \hspace{0.33cm}$[10^{-3},\,10^{3}]$ & Log \\ \hline \hline
		\end{tabular}
		\caption{Ranges and priors for 21 (12 free + 9 nuisance) model parameters.~The parameters $\{\zeta_1,\,\zeta_2\}$ belong to the $R$ matrix of the Casas-Ibarra parametrization~\cite{Casas:2001sr} (see appendix~\ref{app:yuk}), whereas $\alpha$ is the Majorana phase in the PMNS matrix. The terms NO (IO) refer to Normal (Inverted) Ordering, whereas $\Delta m_{3l}^2 \equiv m_3^2 - m_1^2 > 0$ (NO) and $m_3^2 - m_2^2 < 0$ (IO).}
		\label{tab:param} 
	\end{center}
\end{table}

In the next section, we show various two-dimensional (2D) plots of the profile likelihood ratio (PLR) \cite{Cranmer:2006aga} in the relevant parameter planes or key observables of interest.~Model parameters that are not shown in those plots are profiled over, i.e., the composite log-likelihood function in eq.~\eqref{eqn:tot_like} is maximised with respect to those parameters.~Using Wilks' theorem \cite{Wilks:1938dza}, the PLR can be used as a test statistic to approximately construct the $1\sigma$ ($\sim 68.3\%$) and $2\sigma$ ($\sim 95.4\%$) CL contours \cite{Cowan:2010js}.

\section{Results}\label{sec:results}
We start by showing results for the scalar DM in the case of no mixing between the singlet and doublet (e.g., $\theta = 0$, $\pi/2$). For scalar doublet DM, we reproduce the standard results \cite{LopezHonorez:2006gr,Gustafsson:2012aj}; we consider the CP-even scalar, but the results for the CP-odd scalar are similar. For singlet DM, we also reproduce the results from the literature~\cite{Burgess:2000yq,Cline:2013gha,Athron:2018ipf,Athron:2018hpc}.\footnote{There is an extra parameter with respect to both models separately, $\lambda_{\Phi\varphi}$, but as expected, we see that it does not affect the model phenomenology.} 

Next, we turn on the mixing angle $\theta$ and consider separately the case of real scalar $\eta_2$, real pseudoscalar $A$ and Majorana fermion $\psi_2$ as DM candidates.~We pay special attention on studying how the parameter space opens up in each cases with respect to the usual ScM.

\subsection{No mixing case: Scotogenic model + scalar singlet}
From eq.~\eqref{eqn:mixingSC}, the physical state $\eta_2$ in the case of no mixing is 
\begin{equation}
	\eta_2 \equiv
	\begin{cases}
		\textrm{scalar singlet} \, \varphi\,, & \theta = 0\,, \\[1mm]
		\textrm{scalar doublet} \,\, \phi_R\,,  & \theta = \pi/2\,.
	\end{cases}
\end{equation}
In both cases, the trilinear scalar coupling $ \kappa=0 $, and the model reduces to the usual Scotogenic model plus a scalar singlet; we indeed recover the results from the literature. 

\begin{figure}[t]
	\centering
	
	\includegraphics[width=0.49\textwidth]{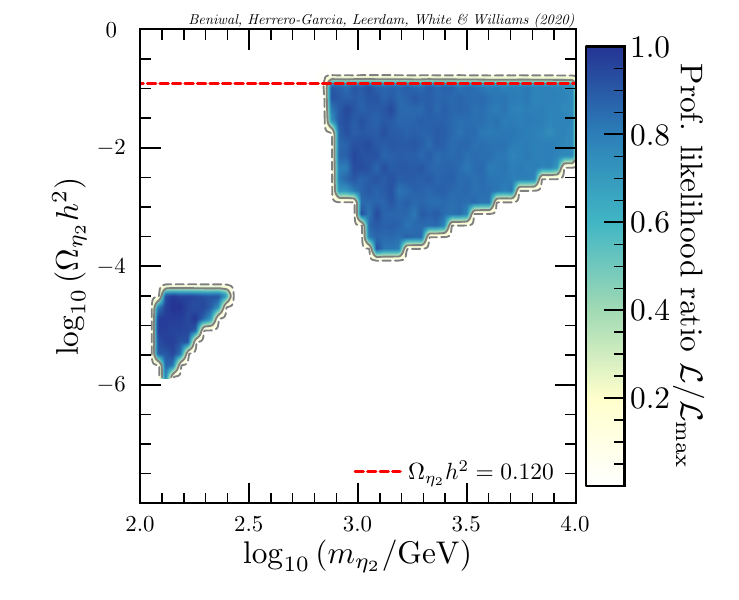}
	\includegraphics[width=0.49\textwidth]{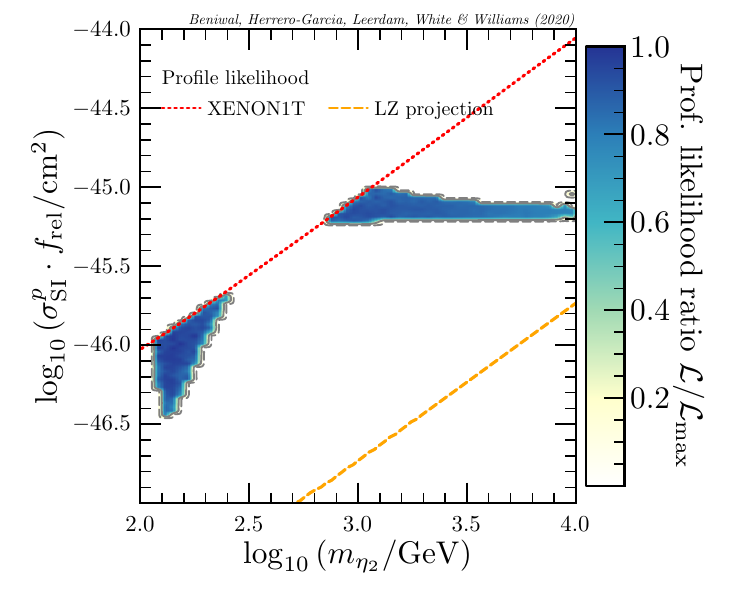}
	
	\caption{2-dimensional (2D) plots of profile likelihood ratio (PLR) for the scalar singlet $\eta_2$ relic density (\emph{left panel}) and spin-independent (SI) $\eta_2$-proton direct detection (DD) cross section (\emph{right panel}), scaled by the DM fraction, $f_{\mathrm{rel}} = \Omega_{\eta_2}/0.120$, for the case of $\theta = 0$ in Normal Ordering (NO).~The dashed (dotted) red lines in the left (right) panel show the \emph{Planck} measured DM abundance \cite{Aghanim:2018eyx} (official XENON1T exclusion limit \cite{Aprile:2018dbl}), whereas the projected LZ sensitivity \cite{Akerib:2018lyp} is shown as dashed orange line in the right panel.}
	\label{fig:singlet}
\end{figure}

In the left (right) panel of figure~\ref{fig:singlet}, we plot the $\eta_2$ relic density (effective SI $\eta_2$ scattering cross section with protons) versus the singlet DM $\eta_2$ mass for $\theta = 0$.~We see two allowed disconnected regions, at masses around $150$ GeV and above $\sim$\,TeV. Notice that only in the latter region, the singlet can constitute $100\%$ of the observed DM abundance \cite{Athron:2018ipf}.~As we see from the plot in the right panel, next-generation DD experiments will be able to test this model, given the fact that the quartic couplings are large enough to reproduce the abundance.

\begin{figure}[t]
	\centering
	
	\includegraphics[width=0.49\textwidth]{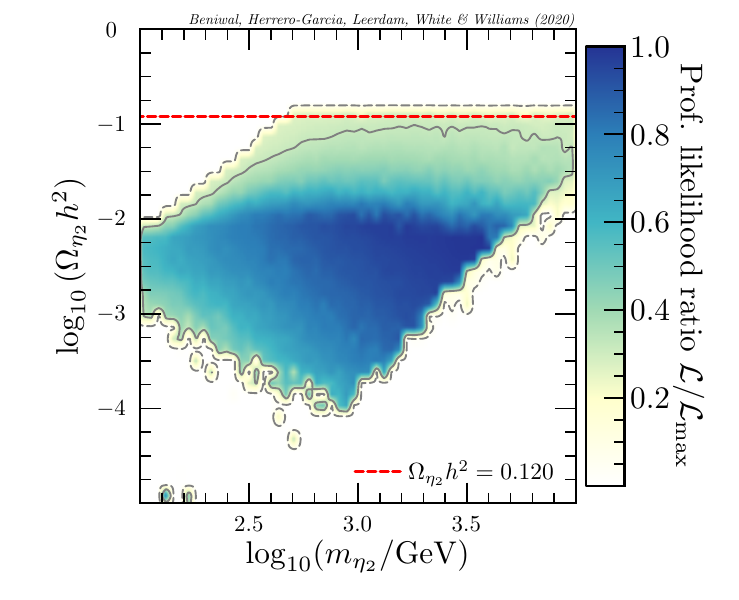}
	\includegraphics[width=0.49\textwidth]{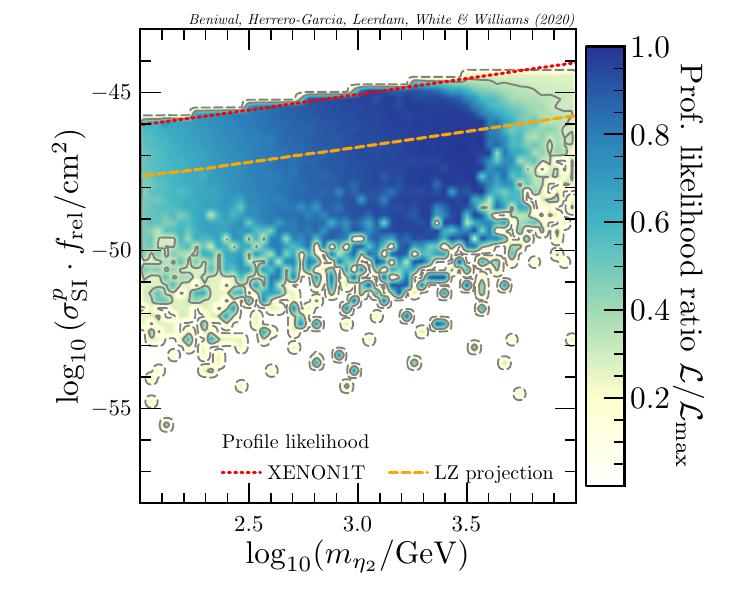}
		
	\caption{Same as figure~\ref{fig:singlet}, but for the case of $\theta = \pi/2$, i.e., pure scalar doublet DM.}
	\label{fig:doublet}
\end{figure}

Similarly in figure~\ref{fig:doublet}, we plot the $\eta_2$ relic density (effective SI $\eta_2$-$p$ cross section) versus the CP-even doublet DM $\eta_2$ mass for $\theta = \pi/2$.\footnote{Notice the under-sampling of the profile likelihood surface at small cross sections. This is expected due to the nature of our XENON1T likelihood (a step-function-like), i.e., parameter points have same likelihood if they are all compatible with the official XENON1T limit. A full coverage of this region requires a dedicated scan over extremely small $h$-$\eta_2$-$\eta_2$ couplings, which is not the main goal of our study.}~The upper left corner in the left plot implies a too-large annihilation cross-section to reproduce the observed DM abundance. Indeed, for the doublet DM to saturate the abundance, its mass should be $\gtrsim 500$ GeV. In this case, next-generation DD experiments will still leave a large portion of the parameter space unexplored. This is due to the fact that the annihilation cross section, driven by gauge interactions, is decoupled from the DD cross section (unless the mass splitting with the CP-odd scalar is $\lesssim$ MeV).

\subsection{DM candidate: real scalar, $\eta_2$}

\begin{figure}[t]
	\centering
	\includegraphics[width=0.49\textwidth]{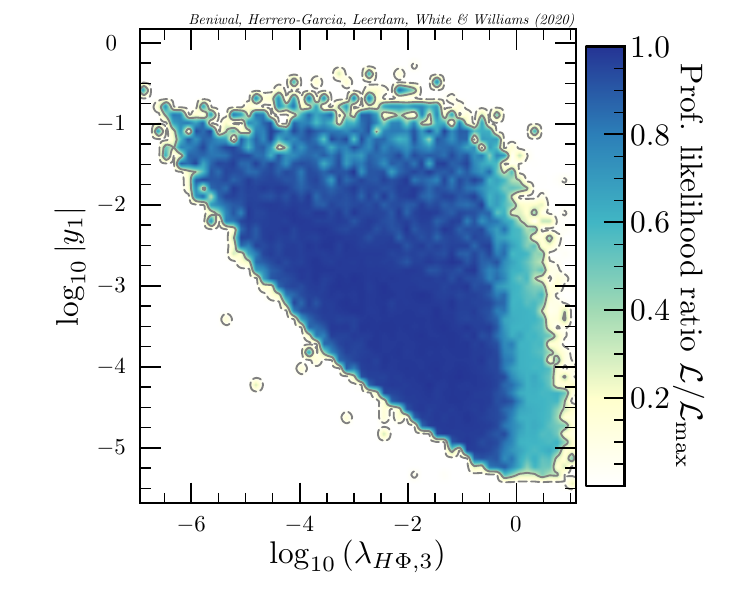}
	\includegraphics[width=0.49\textwidth]{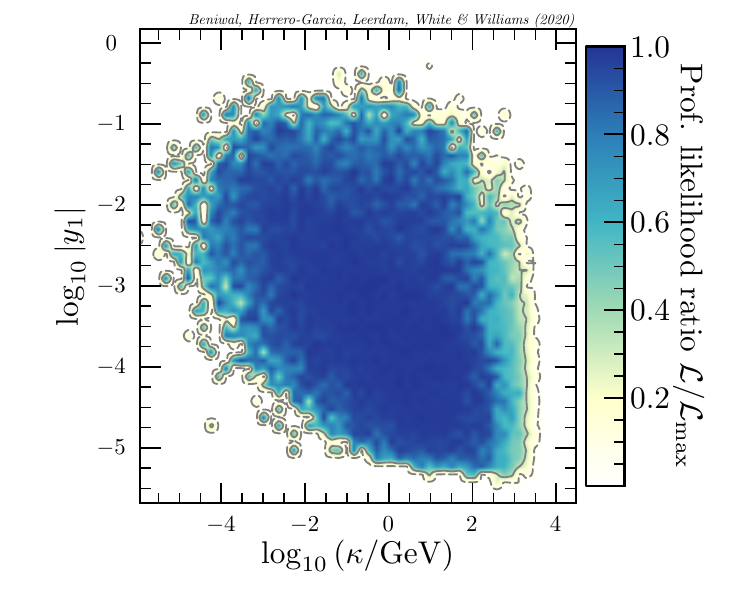}  \\

	\includegraphics[width=0.49\textwidth]{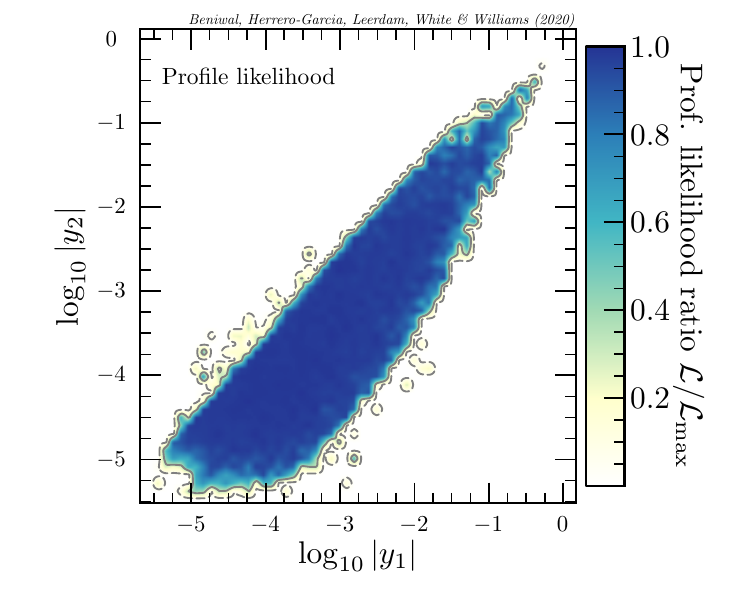}	
	\includegraphics[width=0.49\textwidth]{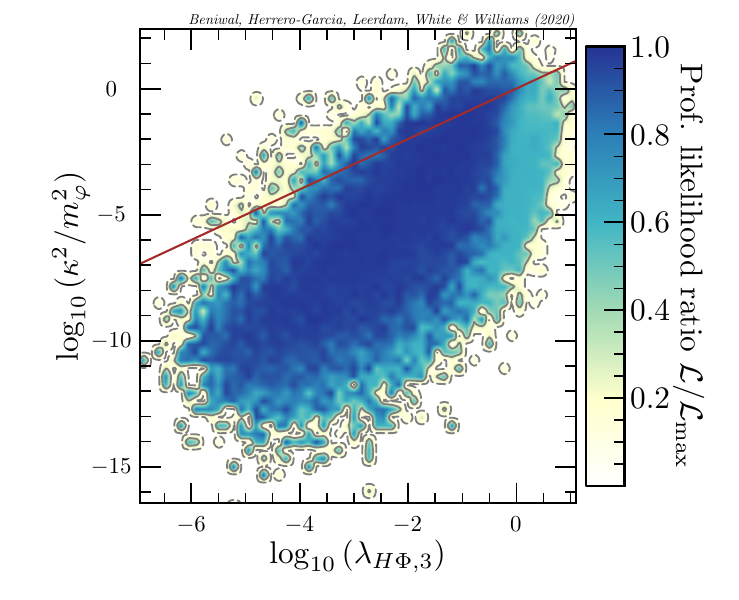} 
	
	\caption{2D plots of PLR in the planes of model parameters for the case of $\eta_2$ DM. In the bottom-right plot, the solid brown line corresponds to a 1:1 relationship between $\kappa^2/m^2_{\varphi}$ and $\lambda_{H\Phi,3}$.}
	\label{fig:eta2-DM-1}
\end{figure}

For the case of CP-even scalar DM $\eta_2$ with non-zero mixing, we find large regions in the model parameter space that can satisfy all included constraints. In figure~\ref{fig:eta2-DM-1}, we plot some of the relevant parameters of interest; Normal ordering (NO) is always assumed, unless stated otherwise. In the top-left plot, we see the expected triangular-shaped region in the Yukawa couplings (plotted for the heaviest sterile singlet fermion $\psi_1$) vs $\lambda_{H\Phi,3}$.~This stems from reproducing the observed neutrino masses.~Small values of $\lambda_{H\Phi,3}$ demand large values for the Yukawa couplings, while for large $\lambda_{H\Phi,3}$ values, large values of the Yukawas are also somewhat possible, and compensated by the masses of the new particles.~A somewhat similar structure is seen in the case of the trilinear coupling $\kappa$ (top-right plot) for the same reasons, although in this case, the region is much less pronounced.~In the bottom-left plot, we see how the Yukawa couplings are correlated amongst each other, with the heaviest sterile ($y_1$) being larger than the lightest one ($y_2$); this correlation becomes even more pronounced at large couplings.~In the bottom-right plot, we observe how the ScSM demands a relationship between the trilinear and quartic couplings, such that neutrino masses are reproduced.~This can be understood analytically in the limit of heavy scalar singlet masses, see eq.~\eqref{eq:L3eff}, as represented by a solid brown line in the plot.

\begin{figure}[t]
	\centering

	\includegraphics[width=0.49\textwidth]{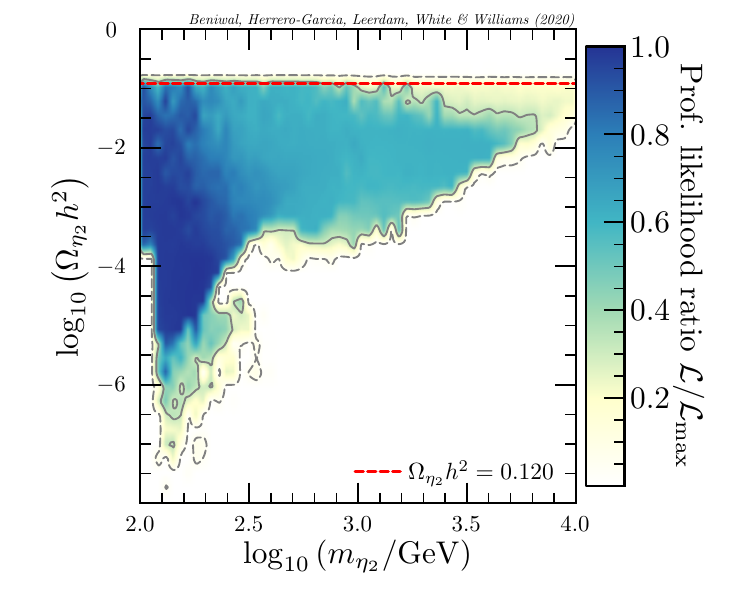}
	\includegraphics[width=0.49\textwidth]{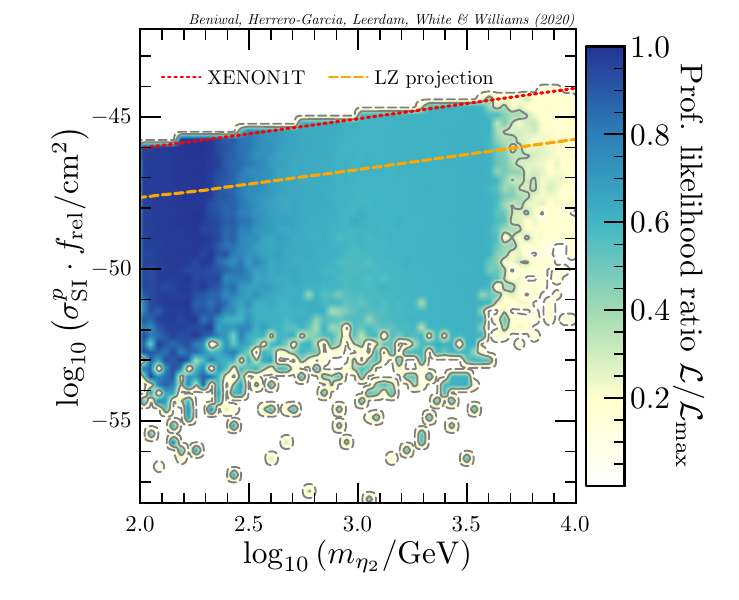} 

	\includegraphics[width=0.49\textwidth]{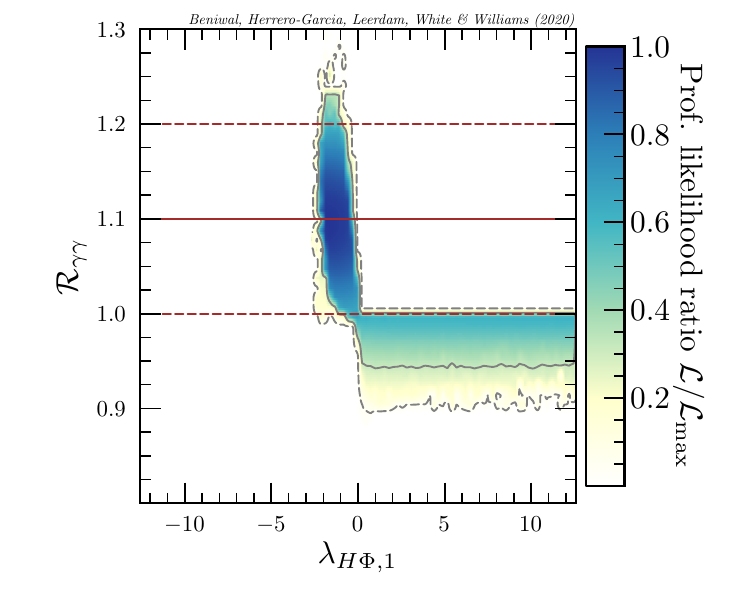}
	\includegraphics[width=0.49\textwidth]{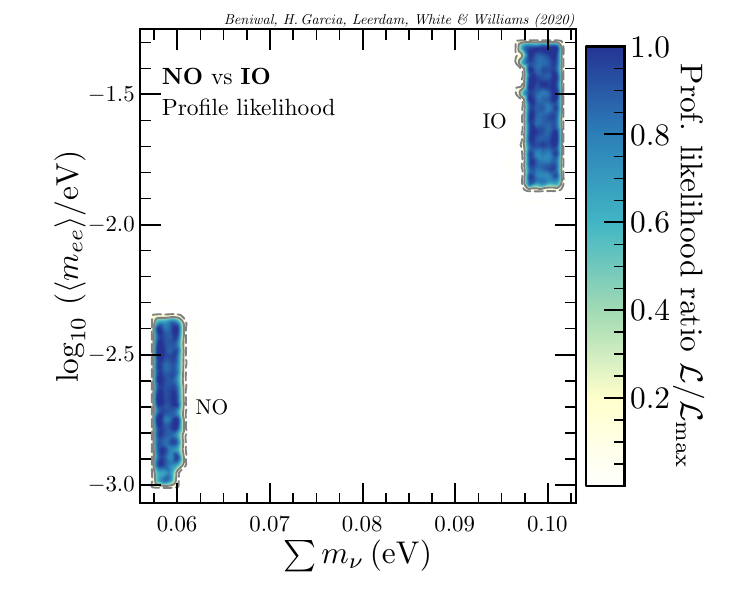}
	
	\caption{2D plots of PLR for key observables of interest.~\emph{Top-left and top-right plots}: same as figure~\ref{fig:singlet} but for $\eta_2$ DM with non-zero mixing.~\emph{Bottom-left plot}: solid (dashed) brown lines show the PDG central (standard deviation) value for $\mathcal{R}_{\gamma \gamma}$ of 1.1 (0.1). \emph{Bottom-right plot}: $\langle m_{ee} \rangle$ vs $\sum m_\nu$ for Normal Ordering (NO) and Inverted Ordering (IO); see text for more details.}
	\label{fig:eta2-DM-2}	
\end{figure}

In the top-left plot in figure~\ref{fig:eta2-DM-2}, the $\eta_2$ abundance is much smaller for $m_{\eta_2} \sim m_h$ due to direct annihilation process $\eta_2 \, \eta_2 \rightarrow h \, h$. Final states with gauge bosons ($W^+W^-/ZZ$) are always open. Indeed, the dominant annihilation channels that determine the $\eta_2$ abundance involve gauge bosons, and less often the Higgs bosons.~Annihilations into $ t \bar t $, leptons or photons are sometimes present.~When the scalar masses are degenerate enough, coannihilations can be important.~An upper limit of $m_{\eta_2} \lesssim 5$ TeV is obtained at the 1$\sigma$ CL, but is somewhat below the upper limit from the prior of 10 TeV. In the top-right plot, we show the effective SI $\eta_2$-proton scattering cross section. The rise in upper limit with DM mass is expected from the DM number density for heavier masses.~The small cross sections arise from a cancellation in eq.~\eqref{eqn:cancel}; in any case, they are well below the sensitivity of next-generation DD experiments.

In the bottom-left plot in figure~\ref{fig:eta2-DM-2}, we observe that the Higgs to di-photon rate with respect to the SM is enhanced (suppressed) for $\lambda_{H\Phi,1}<0$ ($\lambda_{H\Phi,1}>0$). It is evident that the currently allowed value, shown by horizontal brown lines, demands $\lambda_{H\Phi,1}<0$. Finally, in the bottom-right plot, we show the effective neutrino mass parameter $\langle m_{ee} \rangle \equiv \left| \sum U_{ei}^2 \, m_i \right|$, which enters in the expression for the lifetime of neutrinoless double beta ($0\nu\beta\beta$) decay~\cite{Rodejohann:2012xd}. As expected, we reproduce the expected result for NO and IO. In particular, IO results imply that $0.012\,\textrm{eV} \leq \langle m_{ee} \rangle \leq 0.05\,\textrm{eV}$. These values can potentially be tested in coming years, see refs.~\cite{DellOro:2016tmg,Dolinski:2019nrj} for recent reviews.

\begin{figure}[t]
	\centering

	\includegraphics[width=0.49\textwidth]{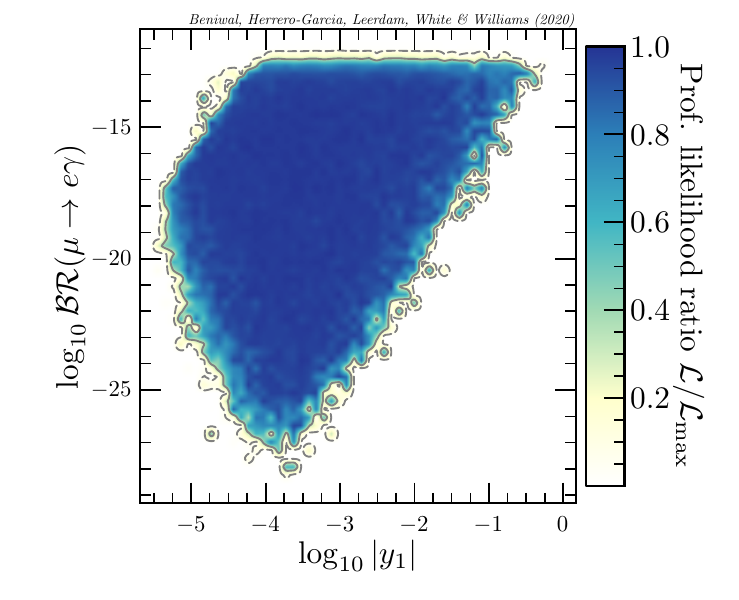}	
	\includegraphics[width=0.49\textwidth]{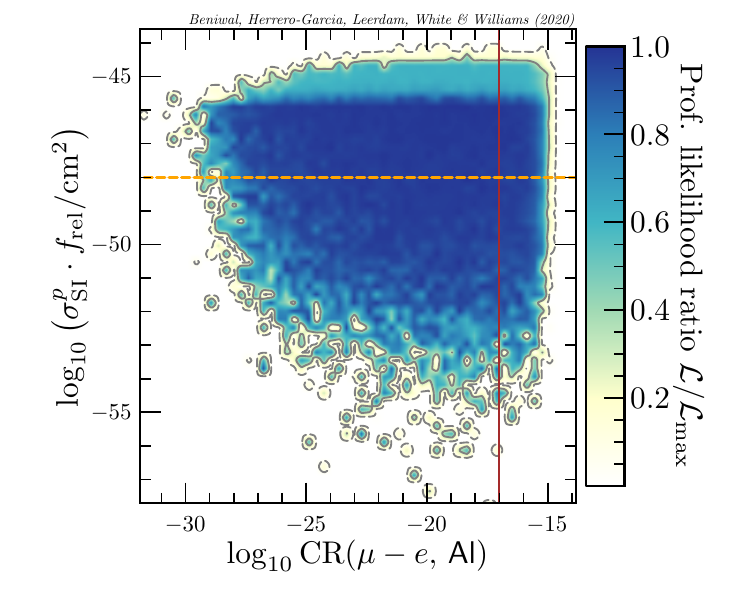}
	
	\caption{\emph{Left panel}:~LFV radiative decay versus the Yukawa of the heaviest fermion singlet.~\emph{Right panel}:~SI DD cross section versus the $\mu \rightarrow e$ conversion rate in aluminium ($\mathsf{Al}$). The solid brown (dashed orange) lines show the expected sensitivity of next-generation $\mu \rightarrow e$ conversion experiments (LZ projection for $50$\,GeV DM \cite{Akerib:2018lyp}).}
	\label{fig:eta2-DM-3}
\end{figure}

In the left panel of figure~\ref{fig:eta2-DM-3}, we plot an LFV radiative decay versus the Yukawa of the heaviest fermion singlet; here we observe a $V$-shaped region.~The behaviour for large Yukawas goes as $\sim |y|^4$, as expected. For $|y_1| \lesssim 10^{-4}$, the contribution from other neutrino ($\propto |y_2|$) dominates.~The upper-left region corresponds to $|y_2| \gg |y_1|$; we see that it is not allowed, as the mass of lightest fermion singlet is too light to suppress enough LFV. In the right panel of figure~\ref{fig:eta2-DM-3}, we see how the next-generation DD experiments (e.g., LZ projected sensitivity for $50$\,GeV DM \cite{Akerib:2018lyp} -- dashed orange line) test complementary parts of the parameter space to those of LFV experiments (e.g., expected sensitivity of $\mu-e$ conversion rate (an improvement by 4 orders of magnitude) -- solid brown line).
 
We have checked that different LFV observables (see table~\ref{tab:BRs_LFV}) show a clear correlation among themselves.~This is expected from the fact that Yukawa couplings are smaller than one, so that the box (dipole) contributions are suppressed (dominant):
\begin{subequations}
	\begin{align}
		\mathcal{BR}(\mu \rightarrow 3e) &\simeq 6 \times 10^{-3} \,\mathcal{BR}(\mu \rightarrow e \gamma)\,, \\
		\textrm{CR} (\mu-e, \, \mathsf{Al}) &\simeq 10^{-2} \, \mathcal{BR}(\mu \rightarrow e \gamma)\,.
	\end{align}
\end{subequations}
It is interesting to highlight that the sensitivity to $\mathcal{BR}(\mu \rightarrow 3e)$ is expected to improve by up to 4 orders of magnitude \cite{Blondel:2013ia}.
These conclusions have already been obtained in the literature, although for somewhat different versions of the model  \cite{Toma:2013zsa,Vicente:2014wga,Hagedorn:2018spx}.~In addition, we have checked that there are no significant differences between the two mass orderings (NO vs IO). 

\begin{figure}[t]
	\centering
	
	\includegraphics[width=0.49\textwidth]{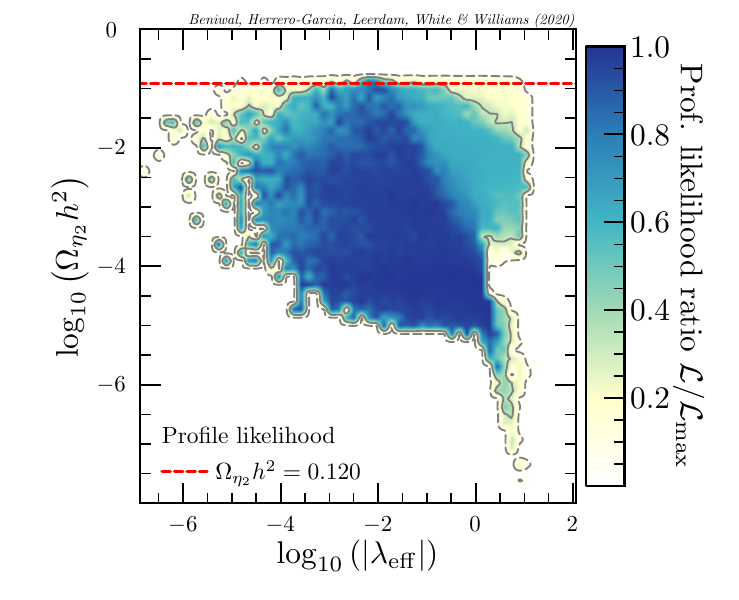}
	\includegraphics[width=0.49\textwidth]{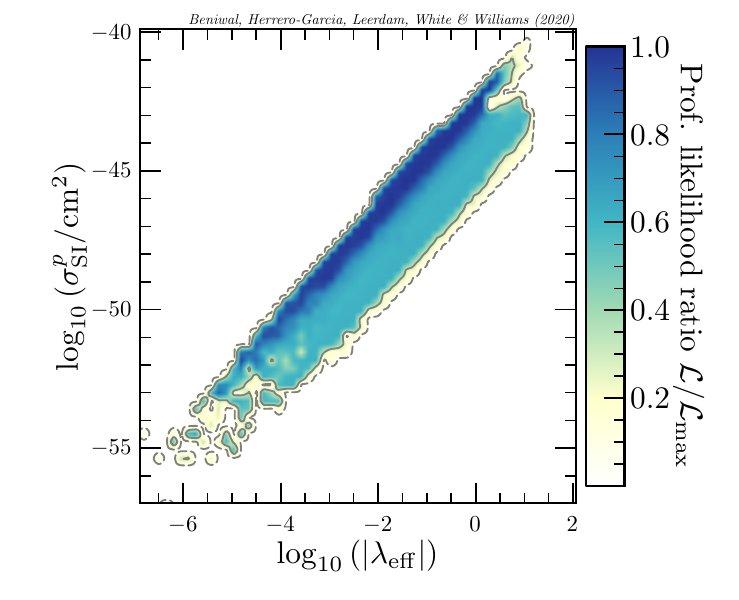}
	
	\caption{2D plots of PLR for the $\eta_2$ relic abundance (\emph{left panel}) and SI $\eta_2$-proton cross section (\emph{right panel}) versus the dimensionless $h$-$\eta_2$-$\eta_2$ coupling, $\lambda_{\mathrm{eff}}$, see eq.~\eqref{eqn:DM-coupling} for more details.}
	\label{fig:xsection_Aeff}
\end{figure}

In the left panel of figure~\ref{fig:xsection_Aeff}, we show how the $\eta_2$ relic abundance changes with the dimensionless $h$-$\eta_2$-$\eta_2$ coupling $\lambda_{\mathrm{eff}}$, see eq.~\eqref{eqn:DM-coupling}.~We observe how the smallest relic abundance is obtained for $\lambda_{\mathrm{eff}}$ values of order one, in which the annihilations proceed via a Higgs-mediated $s$-channel diagram. In the right panel, we see that the SI DD cross section scales linearly with $\lambda_{\mathrm{eff}}$, meaning that $Z$-mediated processes do not contribute significantly to the DD cross section.

\subsection{DM candidate: real pseudoscalar, $A$}
\begin{figure}[t]
	\centering
	
	\includegraphics[width=0.49\textwidth]{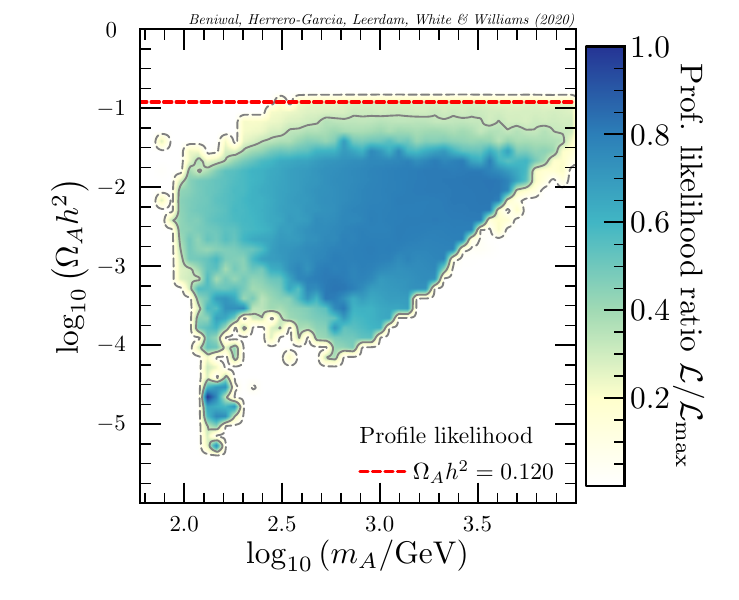}
	\includegraphics[width=0.49\textwidth]{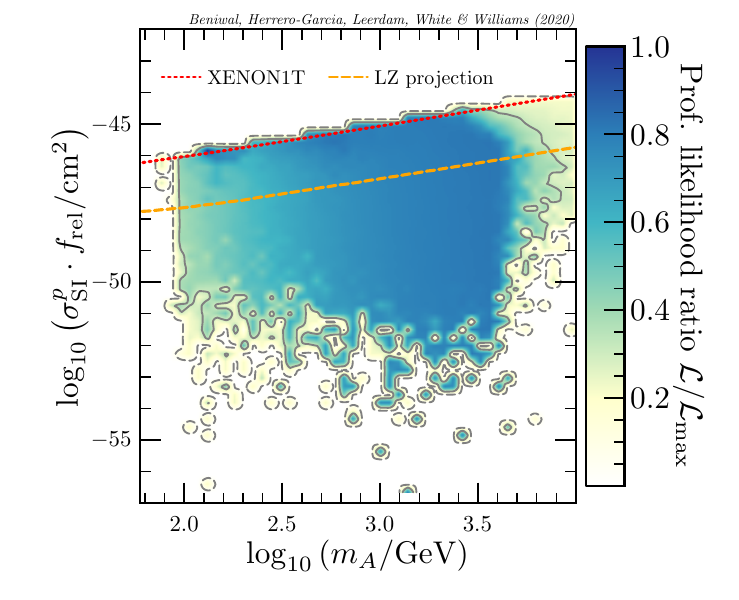}
		
	\caption{Same as figure~\ref{fig:singlet} but for the case of real pseudoscalar $ A $ DM.}
	\label{fig:A-DM}
\end{figure}

The allowed parameter space and phenomenology of the ScSM in this case is similar to the real (CP-even) scalar coming from the doublet (see figure~\ref{fig:doublet}).~The relic abundance of $A$ is also bounded from above for low masses, as shown in the left panel of figure~\ref{fig:A-DM}.~This translates into a lower limit of $m_A \gtrsim 300$\,GeV for pseudoscalar $A$ to saturate the observed DM abundance, somewhat smaller than for the CP-even candidate. This difference comes from the coupling $\lambda_{H \Phi,3}$, which enters differently into the scalar masses. 

\subsection{DM candidate: Majorana fermion, $ \psi_2 $}
\begin{figure}[t]
	\centering
	
	\includegraphics[width=0.49\textwidth]{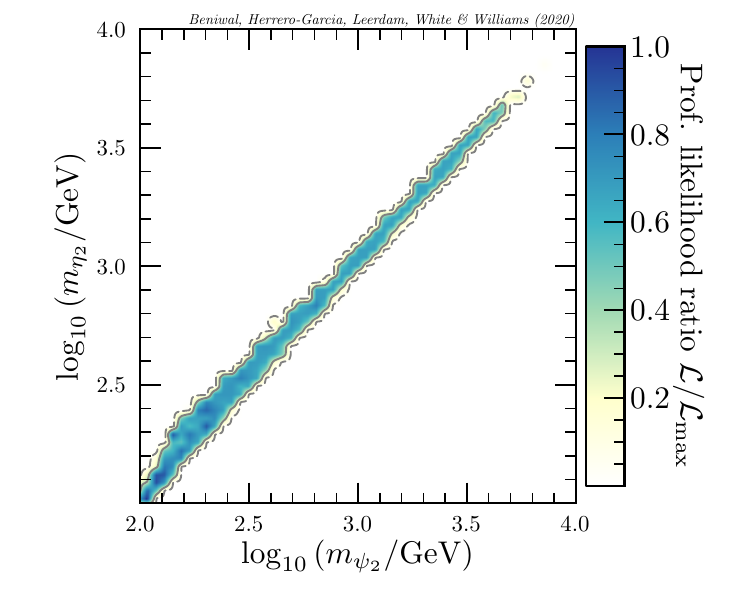}
	\includegraphics[width=0.49\textwidth]{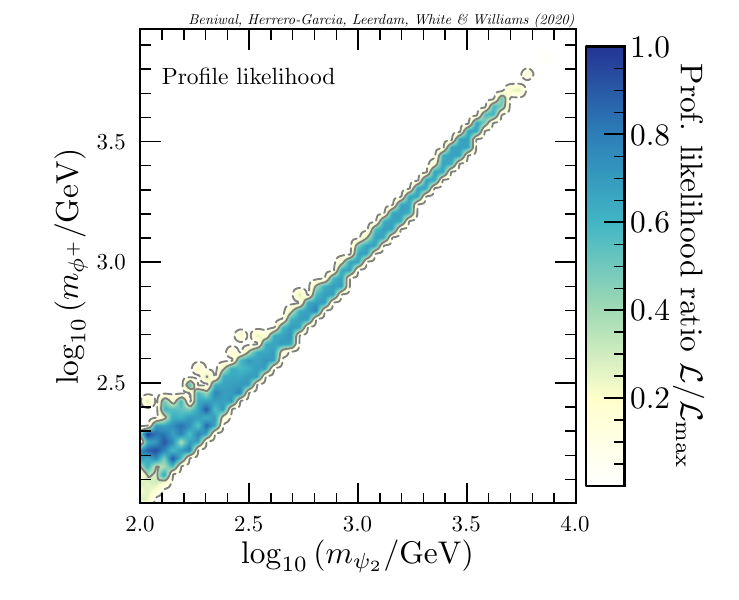}	
	
	\caption{2D plots of PLR in the case of $\psi_2$ DM.~Both panels show the need for a mass degeneracy to saturate the observed DM abundance via coannihilations.}
	\label{fig:psi2-DM}
\end{figure}

We do not investigate the case of fermion DM in detail, as it has already been explored in a model with similar phenomenology \cite{Hagedorn:2018spx}.~Here we simply confirm that we arrive at the same conclusions, namely that the fermion DM case requires coannihilations for its abundance to match the \emph{Planck} measured value.~In figure~\ref{fig:psi2-DM}, we see that viable parameter space requires $m_{\psi_2}$ to be degenerate with $m_{\eta_2}$ (\emph{left panel}) and $m_{\phi^+}$ (\emph{right panel}). This is so because the scalar masses are set by similar combinations of Lagrangian parameters, and EWPT demands them to be close in mass, although there is somewhat a wider region at small masses for the charged scalar. Thus, the case of fermion DM in the ScSM introduces a fine-tuning that is not present in the case of scalar DM. Apart from this difference, the Yukawa couplings and LFV processes are similar to the case of $\eta_2$ and $A$ DM. In regards to DD in the Scotogenic model with a fermion singlet, it occurs at one-loop level, and is typically suppressed \cite{Schmidt:2012yg,Ibarra:2016dlb,Hagedorn:2018spx,Herrero-Garcia:2018koq}.

\section{Conclusions}\label{sec:conc}

We have proposed a simple variation of the original Scotogenic Model (ScM), namely with an extra real scalar singlet. The model, termed the ScotoSinglet Model (ScSM), is arguably the simplest extension of the popular ScM, with a very rich phenomenology and several interesting features: 
\begin{enumerate}
	\item It allows for DM to be scalar (CP even or odd) with a naturally-suppressed direct detection rate, either due to a typically large mass splitting with the opposite-CP scalar (not only dependent on $ \lambda_{H\Phi,3}$ as in ScM), or due to a small mixing, $\theta \simeq 0$. In this case, DM is mainly singlet with a small doublet component, and $Z$-boson mediated interactions are suppressed; 
	
	\item There are two contributions to neutrino masses: the usual Scotogenic one ($\propto \lambda_{H\Phi,3}$) and the new singlet one ($\propto \kappa^2$).~In principle, lepton number violation (e.g., the smallness of neutrino masses) demands both couplings to be small, which is technically natural.~However, in the limit of large singlet mass, a large $\lambda_{H\Phi,3}$ can be cancelled with the trilinear coupling term ($\propto \kappa^2/m^2_\varphi$), see eq.~\eqref{eq:L3eff}.~This allows for the couplings to be larger than usual and contribute to the DM phenomenology, e.g., to DM annihilations and scatterings.
	
	\item The presence of the singlet improves the stability of the $\mathbb{Z}_2$ symmetry up to high-energy scales when the trilinear coupling $\kappa$ is real, as it contributes positively to the evolution of Renormalisation Group Equations (RGEs) for $m^2_\varphi$ and $m^2_\Phi$.
\end{enumerate}

The above features significantly open up the parameter space with respect to the ScM. Extensions to three sterile fermions are not expected to change our results significantly.~Other variations, such a proper RGE study, or a Generalised ScotoSinglet Model~\cite{Hagedorn:2018spx}, is left out for a future work.

The origin of neutrino masses, the nature of DM, and their possible connection remains an open question that may possibly take several decades to fully understand.~While we wait eagerly for a positive signal, the study of simplified models (such as the ScSM) allows us to gain insight into the big puzzles, and search for new correlations among different observables that can help us in distinguishing models among a plethora of possibilities.

\acknowledgments{We thank Arcadi Santamaria and C\'eline Degrande for helpful discussions.~AB is supported by fund for Scientific Research F.N.R.S. through the F.6001.19 convention.~JHG is supported by the Generalitat Valenciana through the GenT Excellence Program (CIDEGENT/2020/020).~NL, MW, and AGW are supported by the Australian Research Council (ARC) Centre of Excellence for Particle Physics at the Terascale (CoEPP) (CE110001104) and the ARC Discovery Project grant DP180102209. 

Computational resources have been provided by the Consortium des \'Equipements de Calcul Intensif (C\'ECI), funded by the Fonds de la Recherche Scientifique de Belgique (F.R.S.-FNRS) under Grant No. 2.5020.11 and by the Walloon Region.~We acknowledge the use of \textsf{pippi v2.0} \cite{Scott:2012qh} for generating our 2-dimensional profile likelihood ratio plots.

\appendix

\section{Mass eigenstate basis}\label{app:massbasis}
For the weak eigenstates $\calA = (\phiR\,,\,\varphi)^T$, the mass-term is given by
\begin{equation}
	\mathscr{L}_{\textrm{mass-term}} = -\frac{1}{2} \calA^T \calM^2 \calA \,,
\end{equation}
where 
\begin{equation}
	\calM^2 
	=
	\begin{pmatrix}
		\dfrac{\partial^2 V}{\partial \phiR^2} & \dfrac{\partial^2 V}{\partial \phiR \, \partial \varphi} \\[4mm]
		\dfrac{\partial^2 V}{\partial \varphi \, \partial \phiR} & \dfrac{\partial^2 V}{\partial \varphi^2}
	\end{pmatrix}
	= 
	\begin{pmatrix}
		a & c \\
		c & b
	\end{pmatrix}
\end{equation}
is a (non-diagonal) squared mass matrix. To diagonalise $\calM^2$, we define the mass eigenstates $(\eta_1,\,\eta_2)$ as 
\begin{equation}
	\begin{pmatrix}
		\eta_1 \\ 
		\eta_2 
	\end{pmatrix}
	 =
	 \begin{pmatrix}
	 	\cos\theta & \sin\theta \\
	 	-\sin\theta & \cos\theta
	 \end{pmatrix}
	 \begin{pmatrix}
	 	\phiR \\
	 	\varphi
	 \end{pmatrix}\,,
\end{equation}
where $\theta$ is the mixing angle. Thus, 
\begin{equation}
	\begin{pmatrix}
	 	\phiR \\
	 	\varphi
	 \end{pmatrix}
	 =
	 \calO
	 \begin{pmatrix}
		\eta_1 \\ 
		\eta_2 
	\end{pmatrix}, 
	\quad
	\calO = 
	\begin{pmatrix}
	 	\cos\theta & -\sin\theta \\
	 	\sin\theta & \cos\theta
	 \end{pmatrix}.
\end{equation}
In the mass eigenstate basis, a squared mass matrix satisfies the following relation:
\begin{equation}
	\calO^T \calM^2 \calO = \calD \equiv 
	\begin{pmatrix}
		m_{\eta_1}^2 & 0 \\
		0 & m_{\eta_2}^2 
	\end{pmatrix}.
\end{equation}
Following the analysis in appendix B of ref.~\cite{Beniwal:2018hyi}, we find that
\begin{subequations}
\begin{align}
	m_{\eta_1}^2 &= a \cos^2\theta + b \sin^2 \theta + c \sin 2\theta\,, \\[2mm]
	m_{\eta_2}^2 &= a \sin^2\theta + b \cos^2 \theta - c \sin 2\theta\,, \\[2mm]
	0 &= - \frac{1}{2} (a-b) \sin 2\theta + c \cos 2\theta\,,
\end{align}
\end{subequations}
where the last equality can also be expressed as
\begin{equation}
	\tan 2\theta = \frac{2c}{a -b}\,.
\end{equation}
In matrix notation, the above expressions read as
\begin{equation}
	\begin{pmatrix}
		m_{\eta_1}^2 \\[2mm]
		m_{\eta_2}^2 \\[2mm]
		0
	\end{pmatrix} 
	= 
	\begin{pmatrix}
		\cos^2 \theta && \sin^2\theta && 2 \sin \theta \cos \theta \\[1mm]
		\sin^2 \theta && \cos^2 \theta && - 2 \sin \theta \cos \theta \\[1mm]
		-\sin\theta \cos\theta && \sin\theta \cos\theta && \cos^2 \theta - \sin^2 \theta 
	\end{pmatrix}
	\begin{pmatrix}
		a \\[2mm]
		b \\[2mm]
		c
	\end{pmatrix}.
\end{equation}
By taking $\theta \rightarrow -\theta$, we can express $(a,\,b,\,c)$ in terms of $( m_{\eta_1}^2,\,m_{\eta_2}^2,\,\theta )$ as
\begin{subequations}
\begin{align}
	a &= m_{\eta_1}^2 \cos^2 \theta + m_{\eta_2}^2 \sin^2 \theta \,, \\[2mm]
	b &= m_{\eta_1}^2 \sin^2 \theta + m_{\eta_2}^2 \cos^2 \theta \,, \\[2mm]
	c &= (m_{\eta_1}^2 - m_{\eta_2}^2) \sin \theta \cos\theta \,.
\end{align}
\end{subequations}
Using the relations for $(a,\,b,\,c)$ from section~\ref{sec:model}, we get
\begin{subequations}
\begin{align}
	m_\Phi^2 &= m_{\eta_1}^2 \cos^2 \theta + m_{\eta_2}^2 \sin^2 \theta - \frac{1}{2} (\lambda_{H\Phi,1} + \lambda_{H\Phi,2} + \textcolor{black}{\lambda_{H\Phi,3}}) v^2\,, \label{eqn:pre1} \\[1.5mm]
	m_{\varphi}^2 &= m_{\eta_1}^2 \sin^2 \theta + m_{\eta_2}^2 \cos^2 \theta - \frac{1}{2} \lambda_{H\varphi} v^2 \,, \\[1.5mm]
	\kappa &= \frac{1}{v} \, (m_{\eta_1}^2 - m_{\eta_2}^2) \sin \theta \cos\theta \,. \label{eqn:pre2}
\end{align}
\end{subequations}

\section{Parameterisation of the Yukawa couplings}\label{app:yuk}
Following the working in the original Casas-Ibarra paper~\cite{Casas:2001sr} (see also ref.~\cite{Lopez-Pavon:2015cga} for a one-loop parametrization, and ref.~\cite{Cordero-Carrion:2019qtu} for a general parametrization), we write the neutrino mass matrix in terms of our high energy parameters, equivalent to eq.~\eqref{neutrinomassmatrix}, as
\begin{align}
	M_\nu = f y^T \hat{M} y\,,
\end{align}
where $ f = 1/(32\pi^2) $, $ y $ is a $ 2\times 3 $ matrix of Yukawa couplings and $ \hat{M} = \mathrm{diag}(\hat{m}_1,\,\hat{m}_2)$ with
\begin{align}
	\hat{m}_k = m_{\psi_k} \Big[ \cos^2\theta\, F_k(\eta_1) + \sin^2\theta \, F_k(\eta_2) - F_k(A) \Big] \,,
\end{align}
where  the loop function is given in eq.~\eqref{eq:loop}. 

The neutrino mass matrix can also be written in terms of the physical neutrino masses ($ m_1 $, $ m_2 $, $ m_3 $) and unitary PMNS matrix $ U$ as
\begin{align}
	U^T M_\nu U = D_\nu \,,
\end{align} 
where $ D_\nu = \text{diag}(m_1,\,m_2,\,m_3) $; in our model with two fermion singlets, $m_1=0$ for NO and $m_3=0$ for IO. The neutrino mass eigenstates $\nu_i$ ($i=1$, $2$, $3$) are related to the neutrino flavour eigenstates $\nu_\alpha$ ($\alpha=e$, $\mu$, $\tau$) by
\begin{equation}
	\nu_\alpha = \sum_{i=1}^3 \,U_{\alpha i} \, \nu_i\,.
\end{equation} 
Now, we can write $ D_\nu = f U^T y^T \hat{M} y U $, and pre- and post-multiply with $ D_{-1/2} $ and $ D_{-1/2}^T $, respectively, to get 
\begin{align}
	I_{2\times 2} = f \, D_{-1/2} \, U^T \, y^T \, \hat{M}^{1/2\, T} \, \hat{M}^{1/2} \, y \, U \, D_{-1/2}^T \,,
\end{align}
where
\begin{align}
	D_{\pm 1/2} \equiv \left\{
	\begin{array}{ll}
	\begin{pmatrix} 0&m_2^{\pm 1/2}&0\\0&0& m_3^{\pm 1/2}	\end{pmatrix},\quad \text{NO}\,, \vspace{0.1cm}\\[6mm]
	\begin{pmatrix} m_1^{\pm 1/2}&0&0\\0&m_2^{\pm 1/2}&0	\end{pmatrix},\quad \text{IO}\,.
	\end{array}
	\right.
\end{align}
We define an orthogonal $ 2\times 2 $ matrix $ R $ as
\begin{align}
	R\equiv \sqrt{f} \, \hat{M}^{1/2} \, y \, U \, D_{-1/2}^T \,,
\end{align}
such that
\begin{align}
	y = \frac{1}{\sqrt{f}} \, \hat{M}^{-1/2} \, R \, D_{1/2} \, U^\dagger\,.
\end{align}

\begin{table}[tb]
	\centering
	\includegraphics[width=\textwidth]{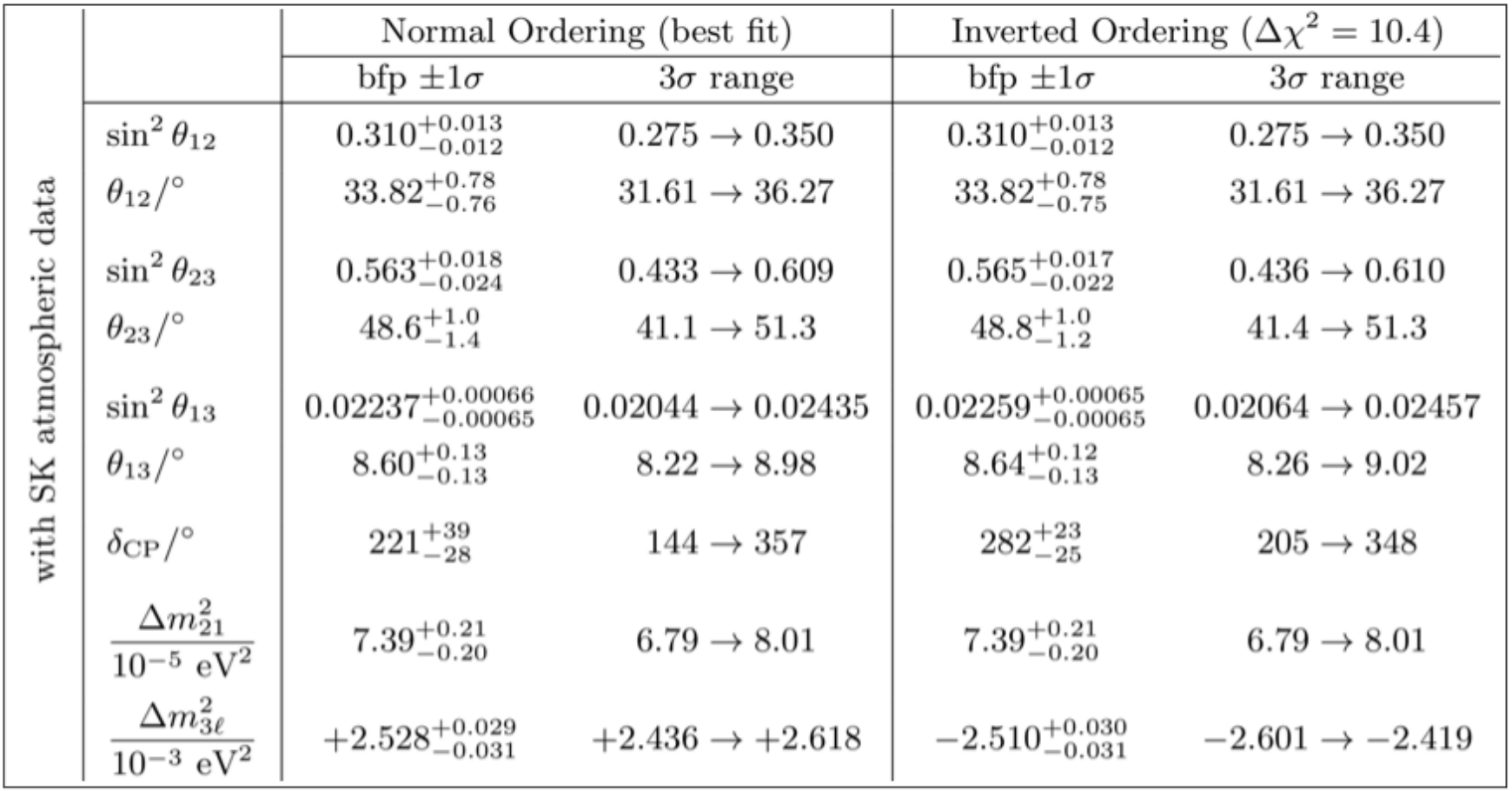}
	\caption{Neutrino oscillation parameters from \textsf{Nu-FIT\,v4.1} (2019) \cite{Esteban:2018azc, nufit2}.}
	\label{tab:nufit}
\end{table}

It is easy to check that this satisfies the low energy definition by substituting back into the high energy expression. We parametrise the $ R $ matrix as
\begin{align}
	R &\equiv \begin{pmatrix} \cos(\zeta_1 + i \zeta_2)&\sin(\zeta_1 +i \zeta_2)\\-\sin(\zeta_1 +i \zeta_2)&\cos(\zeta_1 + i \zeta_2) \end{pmatrix}  \nonumber \\
	&=
	\begin{pmatrix}
		\cos \zeta_1 \cosh \zeta_2 - i \sin \zeta_1 \sinh \zeta_2 & & \sin\zeta_1 \cosh \zeta_2 + i \cos\zeta_1 \sinh\zeta_2 \\
		- \sin\zeta_1 \cosh \zeta_2 - i \cos\zeta_1 \sinh\zeta_2 & & \cos \zeta_1 \cosh \zeta_2 - i \sin \zeta_1 \sinh \zeta_2
	\end{pmatrix}
	 \,,
\end{align}
where $\zeta_1,\,\zeta_2\in \mathbb{R}$. In our numerical scans, we use the standard parametrisation for $U=(u_1,u_2,u_3) $ for one massless neutrino, namely
\begin{equation}
	U=\left(\begin{array}{ccc}
	c_{12}c_{13} & s_{12}c_{13} & s_{13}e^{-i\delta_{\textnormal{CP}}}\\
	-s_{12}c_{23} - c_{12}s_{23}s_{13}e^{i\delta_{\textnormal{CP}}} & c_{12}c_{23} - s_{12}s_{23}s_{13}e^{i\delta_{\textnormal{CP}}} & s_{23}c_{13} \\
	s_{12}s_{23} - c_{12}c_{23}s_{13}e^{i\delta_{\textnormal{CP}}} & -c_{12}s_{23} - s_{12}c_{23}s_{13}e^{i\delta_{\textnormal{CP}}} & c_{23}c_{13}\end{array}\right)\left(\begin{array}{ccc}
	1 & 0 & 0\\
	0 & e^{i\alpha} & 0\\
	0 & 0 & 1\end{array}\right)\,,
	\label{UPMNS}
\end{equation}
where $c_{ij}\equiv\cos\theta_{ij}$ and $s_{ij}\equiv\sin\theta_{ij}$ ($\theta_{12}$, $\theta_{13}$, and $\theta_{23}$ are the 3 lepton mixing angles), $\alpha$ ($\delta_{\textnormal{CP}}$) is the Majorana (Dirac) phase.~As the lightest neutrino is massless with just two fermionic singlets, there is only one physical Majorana phase. For the neutrino oscillation parameters, we use the results based on a global fit from the \textsf{Nu-FIT} collaboration~\cite{Esteban:2018azc, nufit2} with SK atmospheric data (see table~\ref{tab:nufit}).

\section{Potential stability} \label{app:stab}
We follow the analysis in refs.~\cite{Kannike:2012pe,Kannike:2016fmd}.~We parameterise 
\begin{align}
	H^\dagger H = \frac{1}{2} h_1^2,\, \quad \Phi^\dagger \Phi = \frac{1}{2} h_2^2,\, \quad \varphi^2 = h_3^2,\, \quad H^\dagger\Phi = \frac{1}{2} h_1 h_2\rho_{12} e^{i \phi_{12}}\,. 
\end{align}
The quartic part of the potential now reads
\begin{align}
	V_4 &= \lambda_H(H^\dagger H)^2 + \lambda_\Phi(\Phi^\dagger\Phi)^2 + \frac{\lambda_{\varphi}}{4} \varphi^4 + \lambda_{H\Phi,1} (H^\dagger H)(\Phi^\dagger \Phi)  \nonumber \\
	&\hspace{5mm} + \lambda_{H\Phi,2} (H^\dagger \Phi)(\Phi^\dagger H) + \frac{1}{2} \lambda_{H\Phi,3} \big[(H^\dagger\Phi)^2 + \text{H.c.}\big] + \frac{1}{2}\lambda_{H\varphi} (H^\dagger H)\varphi^2 + \frac{1}{2} \lambda_{\Phi\varphi} (\Phi^\dagger \Phi)\varphi^2 \nonumber \\
	&= \begin{pmatrix}
	h_1^2 & h_2^2 & h_3^2
	\end{pmatrix}
	\begin{pmatrix}
	a_{11} & a_{12} & a_{13} \\
	a_{12} & a_{22} & a_{23} \\
	a_{13} & a_{23} & a_{33}
	\end{pmatrix}
	\begin{pmatrix}
	h_1^2 \\ h_2^2 \\ h_3^2
	\end{pmatrix}\,,
\end{align}
where
\begin{subequations}
\begin{align}
	a_{11} = \frac{1}{4} \lambda_H\,, \quad a_{22} &= \frac{1}{4} \lambda_\Phi\,, \quad a_{33} = \frac{1}{4}\lambda_{\varphi}\,, \quad a_{13} = \frac{1}{8}\lambda_{H\varphi}\,, \quad a_{23} = \frac{1}{8} \lambda_{\Phi\varphi}\,, \\[1.5mm]
	a_{12} &= \frac{1}{8} \left[\lambda_{H\Phi,1} + \rho_{12}^2 \Big(\lambda_{H\Phi,2} + \lambda_{H\Phi,3} \cos(2\phi_{12}) \Big)\right]\,.
\end{align}
\end{subequations}
The potential is minimised with respect to $ \rho_{12} $ and $ \phi_{12} $ by setting
\begin{align}
	\cos(2\phi_{12}) = \left\{
	\begin{array}{ll}
		-1\,, & \quad \lambda_{H\Phi,3}>0\,, \\
		1\,,  & \quad \lambda_{H\Phi,3}<0\,, 
	\end{array}
	\right.\qquad \rho_{12} = \left\{
	\begin{array}{ll}
		0\,, & \quad \lambda_{H\Phi,2}- |\lambda_{H\Phi,3}|>0\,, \\
		1\,, & \quad \lambda_{H\Phi,2}- |\lambda_{H\Phi,3}|<0\,. 
	\end{array}
	\right.
\end{align}
The co-positivity conditions are given by
\begin{subequations}
\begin{align}
	\left\{\lambda_{H},\,\lambda_{\Phi},\,\lambda_{\varphi} \right\} &\ge 0\,, \\
	c_1 \equiv \frac{1}{2} \left[\lambda_{H\Phi,1} + \rho_{12}^2(\lambda_{H\Phi,2} - |\lambda_{H\Phi,3}|) \right] + \sqrt{\lambda_{H}\lambda_\Phi} &\ge 0 \,, \\
	c_2 \equiv  \frac{1}{2} \lambda_{H\varphi} + \sqrt{\lambda_{H}\lambda_{\varphi}} &\ge 0 \,, \\
	c_3 \equiv \frac{1}{2}\lambda_{\Phi\varphi} + \sqrt{\lambda_{\Phi}\lambda_{\varphi}} &\ge 0 \,, \\
	\sqrt{\lambda_{H}\lambda_{\Phi} \lambda_{\varphi}} + \frac{1}{2} \Big[\lambda_{H\Phi,1} + \rho_{12}^2(\lambda_{H\Phi,2} - |\lambda_{H\Phi,3}|) \Big] \sqrt{\lambda_{\varphi}} \nonumber \\
	+ \frac{1}{2} \lambda_{H\varphi} \sqrt{\lambda_\Phi} + \frac{1}{2}\lambda_{\Phi\varphi} \sqrt{\lambda_H} + \sqrt{2 c_1 c_2 c_3} &\ge 0\,.
\end{align}
\end{subequations}

\section{Renormalisation Group Equations (RGEs)} \label{app:RGE}
Here we provide the Renormalisation Group Equations (RGEs) for the ScSM at one-loop level, as computed using the \textsf{SARAH} package \cite{Staub:2008uz}.

\subsection{Gauge couplings}

\begin{equation} 
	\beta_{g_1}^{(1)} = \frac{21}{5} g_{1}^{3}\,, \qquad
	\beta_{g_2}^{(1)} = -3 g_{2}^{3}\,, \qquad
	\beta_{g_3}^{(1)} = -7 g_{3}^{3}\,. 
\end{equation}

\subsection{Quartic scalar couplings}
{\allowdisplaybreaks  \begin{align} 
	\beta_{\lambda_H}^{(1)} & =  
	\frac{27}{200} g_{1}^{4} +\frac{9}{20} g_{1}^{2} g_{2}^{2} +\frac{9}{8} g_{2}^{4} -\frac{9}{5} g_{1}^{2} \lambda_H -9 g_{2}^{2} \lambda_H +24 \lambda_{H}^{2} +2 \lambda_{H\Phi,1}^{2} +2 \lambda_{H\Phi,1} \lambda_{H\Phi,2} \nonumber  \\
	&\hspace{5mm} +\lambda_{H\Phi,2}^{2}+\lambda_{H\Phi,3}^{2}+\frac{1}{2} \lambda_{H\varphi}^{2} +12 \lambda_H \mbox{Tr}\Big({Y_d  Y_{d}^{\dagger}}\Big) +4 \lambda_H \mbox{Tr}\Big({Y_e  Y_{e}^{\dagger}}\Big) +12 \lambda_H \mbox{Tr}\Big({Y_u  Y_{u}^{\dagger}}\Big) \nonumber \\
	&\hspace{5mm} -6 \mbox{Tr}\Big({Y_d  Y_{d}^{\dagger}  Y_d  Y_{d}^{\dagger}}\Big) -2 \mbox{Tr}\Big({Y_e  Y_{e}^{\dagger}  Y_e  Y_{e}^{\dagger}}\Big) -6 \mbox{Tr}\Big({Y_u  Y_{u}^{\dagger}  Y_u  Y_{u}^{\dagger}}\Big) \,, \label{lH RGE} \\ 
	\beta_{\lambda_{\Phi}}^{(1)} & =  
	\frac{27}{200} g_{1}^{4} +\frac{9}{20} g_{1}^{2} g_{2}^{2} +\frac{9}{8} g_{2}^{4} +2 \lambda_{H\Phi,1}^{2} +2 \lambda_{H\Phi,1} \lambda_{H\Phi,2} +\lambda_{H\Phi,2}^{2}+\lambda_{H\Phi,3}^{2}-\frac{9}{5} g_{1}^{2} \lambda_{\Phi} \nonumber \\
	&\hspace{5mm} -9 g_{2}^{2} \lambda_{\Phi} +24 \lambda_{\Phi}^{2} +\frac{1}{2} \lambda_{\Phi\varphi}^{2} +4 \lambda_{\Phi} \mbox{Tr}\Big({Y_{\psi}  Y_{\psi}^{\dagger}}\Big) -2 \mbox{Tr}\Big({Y_{\psi}  Y_{\psi}^{\dagger}  Y_{\psi}  Y_{\psi}^{\dagger}}\Big) \,,\\ 
	\beta_{\lambda_{H\Phi,1}}^{(1)} & =  
	\frac{27}{100} g_{1}^{4} -\frac{9}{10} g_{1}^{2} g_{2}^{2} +\frac{9}{4} g_{2}^{4} -\frac{9}{5} g_{1}^{2} \lambda_{H\Phi,1} -9 g_{2}^{2} \lambda_{H\Phi,1} +12 \lambda_H \lambda_{H\Phi,1} +4 \lambda_{H\Phi,1}^{2} \nonumber \\
	&\hspace{5mm} +4 \lambda_H \lambda_{H\Phi,2} +2 \lambda_{H\Phi,2}^{2} +2 \lambda_{H\Phi,3}^{2} +12 \lambda_{H\Phi,1} \lambda_{\Phi} +4 \lambda_{H\Phi,2} \lambda_{\Phi} +\lambda_{H\varphi} \lambda_{\Phi\varphi} \nonumber \\
	&\hspace{5mm} +6 \lambda_{H\Phi,1} \mbox{Tr}\Big({Y_d  Y_{d}^{\dagger}}\Big) +2 \lambda_{H\Phi,1} \mbox{Tr}\Big({Y_e  Y_{e}^{\dagger}}\Big) +2 \lambda_{H\Phi,1} \mbox{Tr}\Big({Y_{\psi}  Y_{\psi}^{\dagger}}\Big) +6 \lambda_{H\Phi,1} \mbox{Tr}\Big({Y_u  Y_{u}^{\dagger}}\Big) \nonumber \\ 
	&\hspace{5mm} -4\, \mbox{Tr}\Big({Y_e  Y_{\psi}^{\dagger}  Y_{\psi}  Y_{e}^{\dagger}}\Big) \,, \\ 
	\beta_{\lambda_{H\Phi,2}}^{(1)} & =  
	\frac{9}{5} g_{1}^{2} g_{2}^{2} -\frac{9}{5} g_{1}^{2} \lambda_{H\Phi,2} -9 g_{2}^{2} \lambda_{H\Phi,2} +4 \lambda_H \lambda_{H\Phi,2} +8 \lambda_{H\Phi,1} \lambda_{H\Phi,2} +4 \lambda_{H\Phi,2}^{2} +8 \lambda_{H\Phi,3}^{2} \nonumber \\
	&\hspace{5mm} +4 \lambda_{H\Phi,2} \lambda_{\Phi} +6 \lambda_{H\Phi,2} \mbox{Tr}\Big({Y_d  Y_{d}^{\dagger}}\Big) +2 \lambda_{H\Phi,2} \mbox{Tr}\Big({Y_e  Y_{e}^{\dagger}}\Big) +2 \lambda_{H\Phi,2} \mbox{Tr}\Big({Y_{\psi}  Y_{\psi}^{\dagger}}\Big) \nonumber \\
	&\hspace{5mm} +6 \lambda_{H\Phi,2} \mbox{Tr}\Big({Y_u  Y_{u}^{\dagger}}\Big) +4 \mbox{Tr}\Big({Y_e  Y_{\psi}^{\dagger}  Y_{\psi}  Y_{e}^{\dagger}}\Big)  \,, \\ 
	\beta_{\lambda_{H\Phi,3}}^{(1)} & =  
	-\frac{9}{5} g_{1}^{2} \lambda_{H\Phi,3} -9 g_{2}^{2} \lambda_{H\Phi,3} +4 \lambda_H \lambda_{H\Phi,3} +8 \lambda_{H\Phi,1} \lambda_{H\Phi,3} +12 \lambda_{H\Phi,2} \lambda_{H\Phi,3} +4 \lambda_{H\Phi,3} \lambda_{\Phi} \nonumber \\
	&\hspace{5mm} +6 \lambda_{H\Phi,3} \mbox{Tr}\Big({Y_d  Y_{d}^{\dagger}}\Big) +2 \lambda_{H\Phi,3} \mbox{Tr}\Big({Y_e  Y_{e}^{\dagger}}\Big) +2 \lambda_{H\Phi,3} \mbox{Tr}\Big({Y_{\psi}  Y_{\psi}^{\dagger}}\Big) \nonumber \\
	&\hspace{5mm} +6 \lambda_{H\Phi,3} \mbox{Tr}\Big({Y_u  Y_{u}^{\dagger}}\Big) \,, \\ 
	\beta_{\lambda_{\varphi}}^{(1)} & =  
	2 \Big(9 \lambda_{\varphi}^{2}  + \lambda_{H\varphi}^{2} + \lambda_{\Phi\varphi}^{2}\Big) \,, \\ 
	\beta_{\lambda_{H\varphi}}^{(1)} & =  
	-\frac{9}{10} g_{1}^{2} \lambda_{H\varphi} -\frac{9}{2} g_{2}^{2} \lambda_{H\varphi} +12 \lambda_H \lambda_{H\varphi} +4 \lambda_{H\varphi}^{2} +4 \lambda_{H\Phi,1} \lambda_{\Phi\varphi} +2 \lambda_{H\Phi,2} \lambda_{\Phi\varphi} +6 \lambda_{H\varphi} \lambda_{\varphi} \nonumber \\
	&\hspace{5mm} +6 \lambda_{H\varphi} \mbox{Tr}\Big({Y_d  Y_{d}^{\dagger}}\Big) +2 \lambda_{H\varphi} \mbox{Tr}\Big({Y_e  Y_{e}^{\dagger}}\Big) +6 \lambda_{H\varphi} \mbox{Tr}\Big({Y_u  Y_{u}^{\dagger}}\Big) \,, \\ 
	\beta_{\lambda_{\Phi\varphi}}^{(1)} & =  
	12 \lambda_{\Phi} \lambda_{\Phi\varphi}  + 2 \lambda_{H\Phi,2} \lambda_{H\varphi}  + 2 \lambda_{\Phi\varphi} \mbox{Tr}\Big({Y_{\psi}  Y_{\psi}^{\dagger}}\Big)  + 4 \lambda_{H\Phi,1} \lambda_{H\varphi}  + 4 \lambda_{\Phi\varphi}^{2}  + 6 \lambda_{\Phi\varphi} \lambda_{\varphi} \nonumber \\
	&\hspace{5mm} -\frac{9}{10} g_{1}^{2} \lambda_{\Phi\varphi}  -\frac{9}{2} g_{2}^{2} \lambda_{\Phi\varphi}\,.
\end{align}}

\subsection{Yukawas, masses and trilinear couplings}
\begin{align} 
	\beta_{Y_{\psi}}^{(1)} & =  
	\frac{1}{2} \Big(3 {Y_{\psi}  Y_{\psi}^{\dagger}  Y_{\psi}}  + {Y_{\psi}  Y_{e}^{\dagger}  Y_e}\Big) + Y_{\psi} \Big[-\frac{9}{20} \Big(5 g_{2}^{2}  + g_{1}^{2}\Big) + \mbox{Tr}\Big({Y_{\psi}  Y_{\psi}^{\dagger}}\Big)\Big]\,, \\
	\beta_{m_{\psi}}^{(1)} & = {m_{\psi}  Y_{\psi}^*  Y_{\psi}^{T}} + {Y_{\psi}  Y_{\psi}^{\dagger}  m_{\psi}}\,, \\
	\beta_{\kappa}^{(1)} & =  
	-\frac{9}{10} g_{1}^{2} \kappa -\frac{9}{2} g_{2}^{2} \kappa +2 \kappa \lambda_{H\Phi,1} +4 \kappa \lambda_{H\Phi,2} +6 \kappa \lambda_{H\Phi,3} +2 \kappa \lambda_{H\varphi} +2 \kappa \lambda_{\Phi\varphi} +3 \kappa \mbox{Tr}\Big({Y_d  Y_{d}^{\dagger}}\Big) \nonumber \\
	&\hspace{4.5mm} +\kappa \mbox{Tr}\Big({Y_e  Y_{e}^{\dagger}}\Big) +\kappa \mbox{Tr}\Big({Y_{\psi}  Y_{\psi}^{\dagger}}\Big) +3 \kappa \mbox{Tr}\Big({Y_u  Y_{u}^{\dagger}}\Big)\,, \\
	\beta_{\mu_H^2}^{(1)} & = -2 \kappa^{2} -\frac{9}{10} g_{1}^{2} \mu_H^2 -\frac{9}{2} g_{2}^{2} \mu_H^2 +12 \lambda_H \mu_H^2 -4 \lambda_{H\Phi,1} m_{\Phi}^2 -2 \lambda_{H\Phi,2} m_{\Phi}^2 -\lambda_{H\varphi} m_{\varphi}^2 \nonumber \\
	&\hspace{4.5mm} 6 \mu_H^2 \mbox{Tr}\Big({Y_d  Y_{d}^{\dagger}}\Big) +2 \mu_H^2 \mbox{Tr}\Big({Y_e  Y_{e}^{\dagger}}\Big) +6 \mu_H^2 \mbox{Tr}\Big({Y_u  Y_{u}^{\dagger}}\Big)\,, \label{mH RGE}\\ 
	\beta_{m_{\Phi}^2}^{(1)} & = 2 \kappa^{2} -4 \lambda_{H\Phi,1} \mu_H^2 -2 \lambda_{H\Phi,2} \mu_H^2 -\frac{9}{10} g_{1}^{2} m_{\Phi}^2 -\frac{9}{2} g_{2}^{2} m_{\Phi}^2 +12 \lambda_{\Phi} m_{\Phi}^2 +\lambda_{\Phi\varphi} m_{\varphi}^2 \nonumber \\
	&\hspace{4.5mm} +2 m_{\Phi}^2 \mbox{Tr}\Big({Y_{\psi}  Y_{\psi}^{\dagger}}\Big) -4 \mbox{Tr}\Big({m_{\psi}  Y_{\psi}  Y_{\psi}^{\dagger}  m_{\psi}}\Big)\,, \label{mPhi RGE}\\ 
	\beta_{m_{\varphi}^2}^{(1)} & =  
	  8 \kappa^{2} -4 \lambda_{H\varphi} \mu_H^2  + 4 \lambda_{\Phi\varphi} m_{\Phi}^2  + 6 \lambda_{\varphi} m_{\varphi}^2\,. \label{mvarphi RGE}
\end{align}
Note that in our convention, $\mu^2_H>0$. As we can see from the last two equations, for real values of $\kappa$, it contributes positively to the running of the bare squared-masses of the $\mathbb{Z}_2$ odd scalars, but negatively to the Higgs doublet bare squared-mass term.

\bibliographystyle{JHEP}
\bibliography{ScotoSinglet}

\end{document}